\documentclass[10pt,a4paper,english,prd,nofootinbib,notitlepage,superscriptaddress,showkeys,preprintnumbers,longbibliography]{revtex4-1}
\usepackage[T1]{fontenc}
\usepackage[latin9]{inputenc}
\usepackage{color}
\usepackage{babel}
\usepackage{amsmath}
\usepackage{amssymb}
\usepackage{graphicx}
\usepackage{esint}
\usepackage{booktabs} 
\usepackage[unicode=true,
 bookmarks=false,
 breaklinks=false,pdfborder={0 0 1},backref=false,colorlinks=true]
 {hyperref}
\hypersetup{
 citecolor=blue,urlcolor=blue,linkcolor=blue}
\usepackage{breakurl}
\usepackage{soul}
\usepackage{todonotes}
\usepackage{multirow}
\usepackage{makecell}

\usepackage{float}

\makeatletter

\usepackage{babel}

\usepackage{babel}
\usepackage{breakurl}\usepackage{cleveref}\usepackage{ulem}\usepackage{bbm}

\usepackage{todonotes}%%
\definecolor{lightblue}{HTML}{A9D0F5}
\definecolor{lightgreen}{HTML}{BCF5A9}
\definecolor{lightred}{HTML}{F6CECE}
\definecolor{lightorange}{HTML}{FFA800}
\definecolor{greengray}{HTML}{5C9393}
\definecolor{lightgreengray}{HTML}{80CCCC}
\makeatother
\begin{document}

%%%%%%%%%%%%%%%%%%%%%%%%%%%%%%%%%%%%%%%%%%%%%%%%%%%%%%%%%%%%%%%%%%%%%%%
\title{Nonlocal Gravity, Restricted CPL Dynamics and DESI Constraints }
%%%%%%%%%%%%%%%%%%%%%%%%%%%%%%%%%%%%%%%%%%%%%%%%%%%%%%%%%%%%%%%%%%%%%%%
%%%%%%%%%%%%%%%%%%%%%%%%%%%%%%%%%%%%%%%%%%%%%%%%%%%%%%%%%%%%%%%%%%%%%%%
\author{Luca Amendola}
\email{l.amendola@thphys.uni-heidelberg.de}
\affiliation{Institut f\"{u}r Theoretische Physik, Ruprecht-Karls-Universit\"{a}t Heidelberg,
Philosophenweg 16, 69120 Heidelberg, Germany}
\affiliation{Center for Astrophysics and Space Science, NYUAD Research Institute, New York University Abu Dhabi, PO Box 129188, Abu Dhabi, UAE}
%%%%%%%%%%%%%%%%%%%%%%%%%%%%%%%%%%%%%%%%%%%%%%%%%%%%%%%%%%%%%%%%%%%%%%%
\author{Ismael Ayuso}
\email{ismael.ayuso@ehu.eus}
\affiliation{Department of Physics \& EHU Quantum Center,
University of the Basque Country UPV/EHU, Bilbao 48080, Spain}
%%%%%%%%%%%%%%%%%%%%%%%%%%%%%%%%%%%%%%%%%%%%%%%%%%%%%%%%%%%%%%%%%%%%%%%
\author{Amlan Chakraborty}
\email{amlan.chakraborty@iiap.res.in}
\affiliation{Indian Institute of Astrophysics, Bengaluru, Karnataka 560034, India}
\affiliation{Institut f\"{u}r Theoretische Physik, Ruprecht-Karls-Universit\"{a}t Heidelberg, Philosophenweg 16, 69120 Heidelberg, Germany}
%%%%%%%%%%%%%%%%%%%%%%%%%%%%%%%%%%%%%%%%%%%%%%%%%%%%%%%%%%%%%%%%%%%%%%%
\author{Javier Rubio}
\email{javier.rubio@ucm.es}
\affiliation{Departamento de F\'isica Te\'orica and Instituto de F\'isica de Part\'iculas y del Cosmos (IPARCOS-UCM), Universidad Complutense de Madrid, 28040 
Madrid, Spain}
%%%%%%%%%%%%%%%%%%%%%%%%%%%%%%%%%%%%%%%%%%%%%%%%%%%%%%%%%%%%%%%%%%%%%%%
%%%%%%%%%%%%%%%%%%%%%%%%%%%%%%%%%%%%%%%%%%%%%%%%%%%%%%%%%%%%%%%%%%%%%%%
\author{Subinoy Das}
\email{subinoy@iiap.res.in}
\affiliation{Indian Institute of Astrophysics, Bengaluru, Karnataka 560034, India}
\affiliation{Institut f\"{u}r Theoretische Physik, Ruprecht-Karls-Universit\"{a}t Heidelberg, Philosophenweg 16, 69120 Heidelberg, Germany}

%%%%%%%%%%%%%%%%%%%%%%%%%%%%%%%%%%%%%%%%%%%%%%%%%%%%%%%%%%%%%%%%%%%%%%%
\begin{abstract}
Nonlocal modifications of gravity provide a theoretically motivated framework in which cosmic acceleration can emerge dynamically without introducing a fundamental cosmological constant. We revisit the cosmological predictions of the Maggiore--Mancarella nonlocal $R\,\Box^{-2}R$ model, with particular emphasis on its background dynamics and observational viability in light of recent DESI baryon acoustic oscillation measurements. A key feature of the theory is its predictive character: for fixed standard cosmological parameters, each value of the nonlocal parameter determines a specific effective pair $(w_0,w_a)$, so that the model traces a restricted trajectory in the Chevallier--Polarski--Linder plane rather than spanning the full phenomenological parameter space. We construct this mapping explicitly and confront the model with observations by implementing its cosmological dynamics in a Boltzmann solver and performing a Markov chain Monte Carlo analysis using Cosmic Microwave Background, supernova, and DESI BAO data. We find that current datasets constrain the nonlocal parameter close to the $\Lambda$CDM limit, with no statistically significant preference for nonlocal deviations.
\end{abstract}
%%%%%%%%%%%%%%%%%%%%%%%%%%%%%%%%%%%%%%%%%%%%%%%%%%%%%%%%%%%%%%%%%%%%%%%

\maketitle

%%%%%%%%%%%%%%%%%%%%%%%%%%%%%%%%%%%%%%%%%%%%%%%%%%%%%%%%%%%%%%%%%%%%%%%
\section{Introduction}\label{sec:intro}
%%%%%%%%%%%%%%%%%%%%%%%%%%%%%%%%%%%%%%%%%%%%%%%%%%%%%%%%%%%%%%%%%%%%%%%

General Relativity (GR) provides an extremely successful description of gravitational phenomena over a vast range of scales, from laboratory and Solar System tests to the large-scale dynamics of the Universe. Combined with the cosmological principle and a cosmological constant, it leads to the standard $\Lambda$CDM model, which provides an excellent fit to a broad set of cosmological observations, including the cosmic microwave background (CMB) \cite{Planck:2018vyg}, large-scale structure, and supernova distance measurements \cite{SupernovaSearchTeam:1998fmf}. Despite this empirical success, the physical origin and the observed small value of the cosmological constant remain theoretically unexplained, giving rise to a severe fine-tuning problem. This has long motivated the exploration of theoretically consistent alternatives in which cosmic acceleration emerges dynamically rather than from a fundamental cosmological constant \cite{Amendola:2015ksp,Bull:2015stt}.

Recent observational developments further reinforce the importance of testing departures from $\Lambda$CDM. In particular, the latest baryon acoustic oscillation measurements from the DESI collaboration have reached unprecedented precision and show a mild preference for dynamical dark-energy parametrizations when combined with other cosmological probes \cite{DESI:2025zgx}. While the statistical significance of these deviations is still under active assessment, see e.g.~\cite{Giare:2025pzu,Capozziello:2025qmh}, they provide a timely opportunity to confront theoretically well-defined models of dynamical dark energy with high-quality data \cite{Ishak:2024jhs,Ye:2024ywg,Pan:2025psn,Wolf:2025acj,Cai:2025mas,Chakraborty:2025syu, Chakraborty:2024xas}. In this context, predictive models, in which the dark-energy evolution is not freely adjustable but instead follows from the underlying theory, are especially valuable. 

An important distinction must therefore be made between phenomenological evidence for an evolving dark-energy component and evidence for a specific theory of cosmic acceleration. While the CPL parametrization allows the present value and time dependence of the dark-energy equation of state to vary independently, a fundamental theory generally correlates these quantities through its equations of motion. Consequently, even a model capable of generating a time-dependent equation of state need not reproduce the particular evolutionary pattern favored by observations. Assessing this compatibility requires confronting the correlations imposed by the underlying dynamics with the data, rather than merely establishing the existence of an evolving dark-energy component.

Among the proposed extensions of GR, nonlocal gravity constitutes a well-motivated class of modified theories \cite{Nojiri:2010wj}. Nonlocal interactions naturally arise in several approaches to quantum gravity. In string theory, for instance, gravitational interactions are expected to acquire a nonlocal character at the Planck scale \cite{Siegel:2003vt,Calcagni:2014vxa}. From an ultraviolet perspective, nonlocality has also been investigated as a possible ingredient for improving the high-energy behavior of gravity and constructing ultraviolet-complete theories \cite{Biswas:2013kla,Biswas:2011ar,Modesto:2011kw}. Nonlocal modifications, however, are not restricted to the ultraviolet regime. They can also emerge at large distances from several infrared mechanisms, including quantum corrections to the effective gravitational action \cite{Woodard:2014iga,Barvinsky:2014lja,Deser:2007jk,Belgacem:2020pdz}, effective field theories obtained after integrating out light degrees of freedom \cite{Codello:2015mba,Donoghue:2015nba}, more fundamental formulations of Euclidean quantum gravity \cite{Wetterich:1997bz,Barvinsky:2011rk}, effective descriptions of massive-graviton theories \cite{Jaccard:2013gla,Modesto:2013jea}, theories involving multiple metrics \cite{Cusin:2014zoa}, and extensions based on post-Riemannian affine geometry \cite{Golovnev:2015bsa}. The phenomenological implications of nonlocal gravity have also been explored in astrophysical settings, including compact stellar objects~\cite{Panotopoulos:2021rih}.

At the level of the gravitational action, nonlocality is typically introduced through curvature invariants involving either infinitely many derivatives, encoded in non-polynomial functions of the d'Alembertian operator \cite{Biswas:2011ar}, or inverse differential operators such as $\Box^{-1}$ \cite{Deser:2007jk} or the inverse conformal anomaly operator \cite{Nersisyan:2016jta}. The latter approach was pioneered in the context of cosmology by Deser and Woodard (DW) \cite{Deser:2007jk}, who proposed one of the first nonlocal models aimed at explaining the late-time cosmic acceleration without explicitly introducing a cosmological constant. The DW model modifies the evolution of cosmological perturbations in a way that was initially considered to be in tension with observations \cite{Dodelson:2013sma}. Subsequent studies, however, showed that this tension could be alleviated \cite{Nersisyan:2017mgj}, although a comprehensive confrontation of the model with observational data remains outstanding.

In this work, we focus on the nonlocal $R\Box^{-2}R$ model introduced by Maggiore and Mancarella (MM) \cite{Maggiore:2014sia}. In this framework, deviations from GR are controlled by a single mass scale and manifest themselves at cosmological scales as an effective dark-energy component. The model modifies the large-scale gravitational dynamics while preserving the successful local predictions of GR and yielding a matter power spectrum compatible with current observations \cite{Dirian:2014xoa,Dirian:2016puz,Dirian:2017pwp}. More generally, the MM model and related realizations of nonlocal gravity have been shown to admit viable cosmological backgrounds and consistent perturbation dynamics without the need for an explicit cosmological constant \cite{Dirian:2014bma,Dirian:2014ara}. An important feature of the MM model is its predictive character: once the realization of the inverse d'Alembertian operator is fixed through the initial conditions of the corresponding auxiliary fields, the cosmological evolution is no longer freely adjustable but is restricted to a narrow family of trajectories controlled by a nonlocal mass parameter \cite{Belgacem:2017cqo}. Consequently, the theory does not span the full space of phenomenological dark-energy parametrizations; instead, its underlying dynamics impose a definite relation between the present-day value of the effective equation of state and its time evolution.

The goal of this paper is to test these theoretically restricted dynamics against current cosmological data, with particular emphasis on the recent DESI results \cite{DESI:2025zgx}. We first review the background evolution of the model using a dynamical-systems approach and identify the class of solutions capable of producing a slowly varying dark-energy component around the present epoch. We then construct a local mapping between these solutions and the effective CPL parameters $(w_0,w_a)$ \cite{Chevallier:2000qy,Linder:2002et}. This mapping allows us to formulate the observational test in a particularly transparent way. Rather than asking whether the model can reproduce isolated CPL histories, we determine the one-dimensional trajectory that it predicts in the $(w_0,w_a)$ plane and examine whether this trajectory intersects the region favored by current data once the standard cosmological parameters are varied simultaneously. We perform this test by implementing the cosmological dynamics in a Boltzmann solver and carrying out a Markov chain Monte Carlo analysis using CMB, supernova, and DESI BAO measurements. We find that varying the standard cosmological parameters does not substantially alter the restricted trajectory predicted by the model, and that current data constrain the nonlocal dynamics close to its $\Lambda$CDM-like limit.

Beyond the numerical likelihood analysis, we show that the so-called path-B dynamics admit a simple local characterization in the CPL plane. Close to the $\Lambda$CDM-like endpoint, the effective parameters satisfy a consistency relation of the form $w_a=-\alpha_{\rm NL}(1+w_0)+\mathcal{O}((1+w_0)^2)$, where the coefficient $\alpha_{\rm NL}$ is determined by the underlying nonlocal dynamics. This relation provides a geometrical interpretation of the observational constraints, but does not by itself establish the viability of the model, since variations of the standard cosmological parameters can, in principle, displace and broaden the theoretical trajectory. Determining whether this freedom is sufficient requires the full likelihood analysis performed below.

The paper is organized as follows. In Sec.~\ref{sec:cosmology} we introduce the nonlocal model and derive the cosmological background equations together with their dynamical-system formulation. In Sec.~\ref{sec:numsol} we present numerical solutions and discuss their effective CPL interpretation. In Sec.~\ref{sec:constraints} we present the observational analysis and parameter constraints. We summarize our conclusions in Sec.~\ref{concl}.

%%%%%%%%%%%%%%%%%%%%%%%%%%%%%%%%%%%%%%%%%%%%%%%%%%%%%%%%%%%%%%%%%%%%%%%
\section{Nonlocal \texorpdfstring{$R\,\Box^{-2}R$}{R box^-2 R} gravity and cosmological background} \label{sec:cosmology}
%%%%%%%%%%%%%%%%%%%%%%%%%%%%%%%%%%%%%%%%%%%%%%%%%%%%%%%%%%%%%%%%%%%%%%%

We consider the MM nonlocal gravity model \cite{Maggiore:2014sia}, defined by the action
\begin{equation}\label{eq:action} S=\frac{M_{\rm Pl}^2}{2}\int d^4x\sqrt{-g}\left[R-\frac{m^2}{6}R\Box^{-2}R\right]+S_m\,, \end{equation}
where $M_{\rm Pl}=2.435\times10^{18}\,\mathrm{GeV}$ is the reduced Planck mass, $g$ is the determinant of the spacetime metric $g_{\mu\nu}$, and $R$ is the Ricci scalar. The parameter $m$ sets the scale of the nonlocal modification and controls the departure from General Relativity, while $S_m$ denotes the matter action, which we assume to be minimally coupled to gravity. The covariant d'Alembertian operator is defined as $\Box\equiv g^{\mu\nu}\nabla_\mu\nabla_\nu$, where $\nabla_\mu$ is the metric-compatible covariant derivative, and $\Box^{-2}\equiv\Box^{-1}\Box^{-1}$. The inverse operator $\Box^{-1}$ is defined through the retarded Green function, thereby ensuring causal evolution \cite{Zhang:2016ykx}.

Varying the nonlocal action~\eqref{eq:action} with respect to the metric yields the modified Einstein equations \cite{Maggiore:2014sia},
\begin{equation}\label{eqmRiBox2R} G^\mu{}_\nu-\frac{m^2}{6}K^\mu{}_\nu=8\pi G\,T^\mu{}_\nu\,, \end{equation}
where the tensor
\begin{equation}\label{eq:Kmunu} K^\mu{}_\nu\equiv2S G^\mu{}_\nu-2\nabla^\mu\partial_\nu S+2\delta^\mu{}_\nu\Box S+\delta^\mu{}_\nu\partial_\rho S\partial^\rho U-\frac{1}{2}\delta^\mu{}_\nu U^2-\left(\partial^\mu S\partial_\nu U+\partial_\nu S\partial^\mu U\right) \end{equation}
encodes the contribution of the nonlocal sector. Here, $U$ and $S$ are auxiliary scalar fields defined by
\begin{equation}\label{eq:udefin} \Box U=-R\,, \qquad \Box S=-U\,, \end{equation}
so that $U=-\Box^{-1}R$ and $S=\Box^{-2}R$ for a specified definition of the inverse d'Alembertian. Their general solutions can be written as
\begin{equation}\label{US-eq} U=U_{\rm hom}-\Box_{\rm ret}^{-1}R\,, \qquad S=S_{\rm hom}-\Box_{\rm ret}^{-1}U\,, \end{equation}
where the homogeneous components satisfy
\begin{equation}\label{eqhom} \Box U_{\rm hom}=0\,, \qquad \Box S_{\rm hom}=0\,, \end{equation}
and $\Box_{\rm ret}^{-1}$ denotes the inverse d'Alembertian defined using the retarded Green function \cite{Zhang:2016ykx}.

Although the introduction of $U$ and $S$ renders the equations of motion formally local, the resulting system is not equivalent to an unconstrained local scalar--tensor theory. In particular, the auxiliary fields do not represent additional fundamental propagating degrees of freedom. Their homogeneous contributions are fixed once the nonlocal prescription, including the Green function and the associated initial or boundary conditions, has been fully specified \cite{Maggiore:2014sia}. Different admissible prescriptions may therefore define distinct realizations of the effective nonlocal dynamics, but the auxiliary fields do not introduce independent physical degrees of freedom within a given realization \cite{Maggiore:2014sia}. In the following, we restrict attention to the realization defined by the retarded prescription and the auxiliary-field initial conditions specified below. The observational constraints derived in Sec.~\ref{sec:constraints} consequently apply to this particular realization.

To study the cosmological implications of the model, we assume a spatially flat Friedmann--Lema\^itre--Robertson--Walker (FLRW) spacetime, $ds^{2}=-dt^{2}+a^{2}(t)\,d\vec{x}^{\,2}$, where $a(t)$ is the cosmological scale factor. The present-day Universe is characterized by the density parameters $\Omega_{m}^{0}$ and $\Omega_{r}^{0}$ for matter and radiation, respectively. The background evolution equations can then be written as \cite{Maggiore:2014sia}
\begin{eqnarray}
h^{2} & = & \dfrac{\Omega_{m}^{0} e^{-3N} + \Omega_{r}^{0} e^{-4N} + (\gamma/4)\,U^{2}}
{1+\gamma\left(-3V - 3V' + \tfrac{1}{2} V'U'\right)},
\label{eq:h2ithomega}
\\
U'' & = & 6\left(2+\xi\right) - \left(3+\xi\right) U',
\label{eq:u1}
\\
V'' & = & h^{-2} U - \left(3+\xi\right) V',
\label{eq:v1}
\end{eqnarray}
where $U$ and $V\equiv H_{0}^{2}S$ are the auxiliary fields introduced in the local formulation of the theory, and
\begin{equation}
\gamma \equiv \frac{m^{2}}{9H_{0}^{2}},
\qquad
h \equiv \frac{H}{H_{0}},
\qquad
\xi \equiv \frac{h'}{h},
\label{eq:Y}
\end{equation}
are dimensionless variables defined in terms of the present-day Hubble rate $H_{0}$. A prime denotes differentiation with respect to the number of $e$-folds, $N\equiv\ln a$. Note that, in the formal limit $\gamma\to0$, the nonlocal contribution vanishes and the equations reduce to those of General Relativity without a cosmological constant. However, once the normalization condition $H(N=0)=H_{0}$ is imposed, the required initial value of the auxiliary field $U$ diverges as $\gamma\to0$, while the corresponding product $\gamma U^{2}$ remains finite, so that the background expansion approaches that of $\Lambda$CDM. This singular scaling limit therefore reproduces a $\Lambda$CDM-like background evolution despite the absence of an explicit cosmological constant term in the action.

The background expansion is conveniently characterized by the effective total equation-of-state parameter,
\begin{align}
w_{\rm eff}=-1-\frac{2}{3}\frac{h'}{h}=-1-\frac{2}{3}\,\xi\,.
\label{eq:weffdef-1}
\end{align}
The matter and radiation components are conserved separately. Combined with the Bianchi identity, this also implies a conservation equation for the effective dark-energy sector, such that 
\begin{equation}
\Omega'_{\rm m}+\left(3+2\xi\right)\Omega_{m}=0,\qquad\quad 
\Omega'_{\rm r}+\left(4+2\xi\right)\Omega_{r}=0,\qquad\quad 
\Omega'_{\rm DE}+\left(3+3w_{\rm DE}+2\xi\right)\Omega_{\rm DE}=0,
\label{eq:conservation}
\end{equation}
where
\begin{equation}
\xi=\frac{-4\Omega_{r}-3\Omega_{m}+3\gamma\left(h^{-2}U+U'V'-4V'\right)}{2\left(1-3\gamma V\right)}.
\label{eq:dittadef}
\end{equation}
Using the cosmic sum rule together with the modified Friedmann equation, the effective dark-energy density fraction may be expressed equivalently as
\begin{equation}
\Omega_{\rm DE}
=1-h^{-2}\left(\Omega_{m}^{0}e^{-3N}+\Omega_{r}^{0}e^{-4N}\right)
=\gamma\left(\frac{1}{4}h^{-2}U^{2}+3V+3V'-\frac{1}{2}V'U'\right).
\label{eq:OmegaDE}
\end{equation}
The corresponding effective dark-energy equation of state is
\begin{equation}
w_{\rm DE}=
\frac{
\gamma\left(4(U+3)-(U+2)U'\right)V'
+U\left(\gamma V(\Omega_{r}+3)-\Omega_{\rm DE}\right)
+4\left(3\gamma V-\Omega_{\rm DE}\right)
}{
U(1-3\gamma V)\Omega_{\rm DE}
}.
\label{eq:wDE}
\end{equation}

The background equations derived above constitute a coupled nonlinear system whose global behavior is not immediately transparent. To elucidate the structure of its cosmological solutions and identify the main dynamical regimes, we recast the equations as an autonomous system. This formulation allows the critical points to be identified, their stability to be assessed, and the admissible cosmological trajectories to be classified within a unified phase-space description.

Introducing the auxiliary variables $Y_1\equiv U'$ and $Y_2\equiv V'$, the dynamics can be written as a first-order system in the six-dimensional phase space $(U,V,Y_1,Y_2,\Omega_{m},\Omega_{r})$. Combining the auxiliary-field equations with the conservation equations for matter and radiation yields a closed autonomous system
\begin{eqnarray}
U' & = & Y_1,
\label{eq:auto1}\\
V' & = & Y_2,\\
Y'_{1} & = & -\frac{3((U+2)Y_{1}-4(U+3))(2(3\gamma V-1)-\gamma(Y_{1}-6)Y_{2})+3(U+4)(Y_{1}-6)\Omega_{m}+4(U+3)(Y_{1}-6)\Omega_{r}}{2U(3\gamma V-1)},
\label{eq:auto2}\\
Y'_{2} & = & -\frac{Y_{2}(6(3\gamma V-1)-3\gamma(Y_{1}-4)Y_{2}+3\Omega_{m}+4\Omega_{r})}{2(3\gamma V-1)}
\nonumber\\
& - &
\dfrac{(2(3\gamma V-1)+3\gamma Y_{2})(2(3\gamma V-1)-\gamma(Y_{1}-6)Y_{2}+2(\Omega_{m}+\Omega_{r}))}{\gamma U(3\gamma V-1)},
\\
\Omega'_{\rm m} & = &
-\frac{\Omega_{m}\big(U(3(3\gamma V-1)-3\gamma(Y_{1}-4)Y_{2}+3\Omega_{m}+4\Omega_{r})+12(3\gamma V-1)+12(\Omega_{m}+\Omega_{r})-6\gamma(Y_{1}-6)Y_{2}\big)}{U(3\gamma V-1)},
\\
\Omega'_{\rm r} & = &
-\frac{\Omega_{r}\big(U(4(3\gamma V-1)-3\gamma(Y_{1}-4)Y_{2}+3\Omega_{m}+4\Omega_{r})+12(3\gamma V-1)+12(\Omega_{m}+\Omega_{r})-6\gamma(Y_{1}-6)Y_{2}\big)}{U(3\gamma V-1)}.
\label{eq:auto_last}
\end{eqnarray}
The asymptotic solutions of this system describe cosmological regimes characterized by a constant effective equation-of-state parameter, although some phase-space variables may diverge in the corresponding limits. Their existence and stability were analyzed in detail in Ref.~\cite{Nersisyan:2016hjh}. The phase space contains radiation- and matter-dominated saddle points, together with accelerated late-time solutions that may behave either as attractors or as transient saddle configurations. The relevant critical points and their main properties are summarized in Table~\ref{tablepoints}.

%%%%%%%%%%%%%%%%%%%%%%%%%%%%%%%%%%%%%%%%%%%%%%%%%%%%%%%%%%%%%%%%%%%%%%%
\begin{table}[t]
\centering
\begin{tabular}{c c c c c c c c c}
\toprule
Point & $U$ & $V$ & $U'$ & $V'$ & $\Omega_{m}$ & $\Omega_{r}$ & $w_{\rm eff}$ & Type \\
\midrule
I   & $\widetilde U$ & $\dfrac{1-\widetilde\Omega_{r}}{3\gamma}$ 
    & $0$ & $0$ 
    & $0$ & $\widetilde\Omega_{r}$ 
    & $\tfrac{1}{3}$ & Saddle \\

II  & $2N+\widetilde U$ & $\dfrac{1-\widetilde\Omega_{m}}{3\gamma}$ 
    & $2$ & $0$ 
    & $\widetilde\Omega_{m}$ & $0$ 
    & $0$ & Saddle \\

III & $+\infty$ & $\dfrac{1}{3\gamma}$ 
    & $4$ & $0$ 
    & $0$ & $0$ 
    & $-1$ & Attractor \\

IV  & $4N+\widetilde U$ & $\pm\infty$ 
    & $4$ & $\pm\infty$ 
    & $0$ & $0$ 
    & $-1$ & Saddle \\

V   & $-3$ & $\pm\infty$ 
    & $0$ & $4V$ 
    & $\mp\infty$ & $0$ 
    & $\tfrac{1}{3}$ & Attractor \\

\bottomrule
\end{tabular}
\caption{Critical points of the autonomous system defined by Eqs.~\eqref{eq:auto1}--\eqref{eq:auto_last}. Tilded quantities denote constant values associated with the corresponding critical point.}
\label{tablepoints}
\end{table}
%%%%%%%%%%%%%%%%%%%%%%%%%%%%%%%%%%%%%%%%%%%%%%%%%%%%%%%%%%%%%%%%%%%%%%%

A key structural feature of the system is the existence of two qualitatively distinct classes of cosmological trajectories, distinguished by whether the initial value of the auxiliary field $U$ lies above or below a critical threshold $\bar U(\gamma)$ \cite{Nersisyan:2016hjh}. Initial conditions with $U_0>\bar U$ define path A \cite{Dirian:2014bma}, whereas those with $U_0<\bar U$ define path B \cite{Nersisyan:2016hjh}. The two branches differ both in their asymptotic behavior and in the sequence of cosmological regimes encountered along their evolution. While path A follows the sequence $\text{I}\rightarrow\text{II}\rightarrow\text{III}$, path B follows the sequence $\text{I}\rightarrow\text{II}\rightarrow\text{IV}\rightarrow\text{V}$. For $\Omega_{m}^{0}=0.3$, the critical value separating them was determined numerically in Ref.~\cite{Nersisyan:2016hjh} as
\begin{equation}
\bar{U}(\gamma)\simeq -14.82 + 0.67 \log\gamma.
\label{eq:ugamma}
\end{equation}
Trajectories along path A typically approach the late-time attractor III and may undergo a transient phantom phase with $w_{\rm eff}<-1$. By contrast, path B trajectories pass near the quasi-de Sitter saddle point, allowing $w_{\rm eff}$ to remain close to $-1$ for an extended interval before the system eventually exits the accelerated regime.

The distinction between these two branches is particularly relevant for late-time cosmology. Although both reproduce the conventional radiation- and matter-dominated stages, their subsequent evolution reflects different ways in which the nonlocal sector can drive cosmic acceleration. The asymptotic classification of the dynamical system therefore provides a useful starting point, but does not by itself determine which branch can reproduce the expansion history inferred from observations. Addressing this question requires examining the evolution near the present epoch, where the effective dark-energy contribution becomes observationally relevant.

%%%%%%%%%%%%%%%%%%%%%%%%%%%%%%%%%%%%%%%%%%%%%%%%%%%%%%%%%%%%%%%%%%%%%%%
\section{Numerical background solutions and CPL mapping}\label{sec:numsol}
%%%%%%%%%%%%%%%%%%%%%%%%%%%%%%%%%%%%%%%%%%%%%%%%%%%%%%%%%%%%%%%%%%%%%%%

Having established the general structure of the dynamical system, we now investigate the background evolution numerically and relate it to the phenomenological CPL description commonly adopted in observational analyses \cite{Chevallier:2000qy,Linder:2002et}. Our aim is to determine how the expansion histories predicted by the nonlocal model map onto the CPL form and to assess the extent to which the theory spans, or instead restricts, the corresponding parameter space.

\begin{figure}
\begin{centering}
\includegraphics[height=8cm]{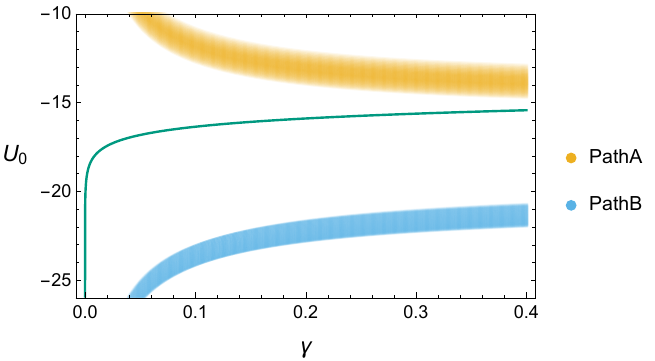}
\end{centering}
\caption{Initial values $U_0$ satisfying the normalization condition $H(N=0)=H_0$, shown as functions of $(\gamma,\Omega_{m}^0,H_0)$. The two solution branches are clearly separated, so the fitting procedure cannot continuously move from one path to the other. The green line corresponds to Eq.~\eqref{eq:ugamma} and marks the approximate boundary between paths A and B for fixed values of $\Omega_{m}^0$ and $H_0$. For each value of $\gamma$ and each branch, the spread of points reflects variations in $\Omega_{m}^0\in[0.25,0.35]$ and $H_0\in[60,80]$.}
\label{paths}
\end{figure}

For fixed values of the cosmological parameters $(\Omega_{m}^0,\Omega_{r}^0,H_0)$ and the model parameter $\gamma$, the background evolution is obtained by numerically integrating the system of equations derived in  Sec.~\ref{sec:cosmology}. For simplicity, we set $V_0=0$ and $V'_0=U'_0=0$ at $N=-20.7233$ $e$-folds before the current epoch, corresponding to a scale factor $a=10^{-9}$ well inside the radiation domination era. In path A, $V_0=0$ can be chosen without loss of generality due to the attractor nature of point III, while in path B this choice remains valid provided that $V_0<1/(3\gamma)-V'_0$, ensuring a physically viable evolution~\cite{Nersisyan:2016hjh}. The initial value $U_0$ is not fully free, as it is constrained by the requirement that the present expansion rate satisfies the normalization condition $H(N=0)=H_0$. In practice, this condition is enforced through a shooting procedure that determines the required initial value $U_0$ for each choice of $\gamma$ and cosmological parameters. Owing to the existence of two possible evolutionary paths,  the shooting procedure generally yields two distinct solutions for $U_0$ corresponding to the same set of cosmological and model parameters. In Fig.~\ref{paths}, we show the numerical values of $U_0$ for path A (yellow points) and path B (cyan points). The figure shows that $U_0$ depends predominantly on $\gamma$, while its dependence on $\Omega_{m}^0$ and $H_0$ is considerably weaker.

%%%%%%%%%%%%%%%%%%%%%%%%%%%%%%%%%%%%%%%%%%%%%%%%%%%%%%%%%%%%%%%%%%%%%%%
%%%%%%%%%%%%%%%%%%%%%%%%%%%%%%%%%%%%%%%%%%%%%%%%%%%%%%%%%%%%%%%%%%%%%%%

\begin{figure}
\begin{centering}
\includegraphics[height=5.5cm]{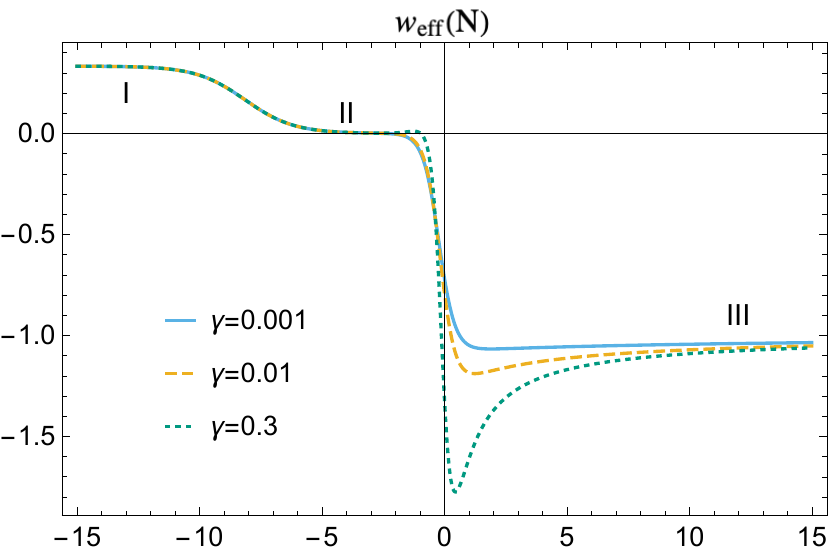}
\includegraphics[height=5.5cm]{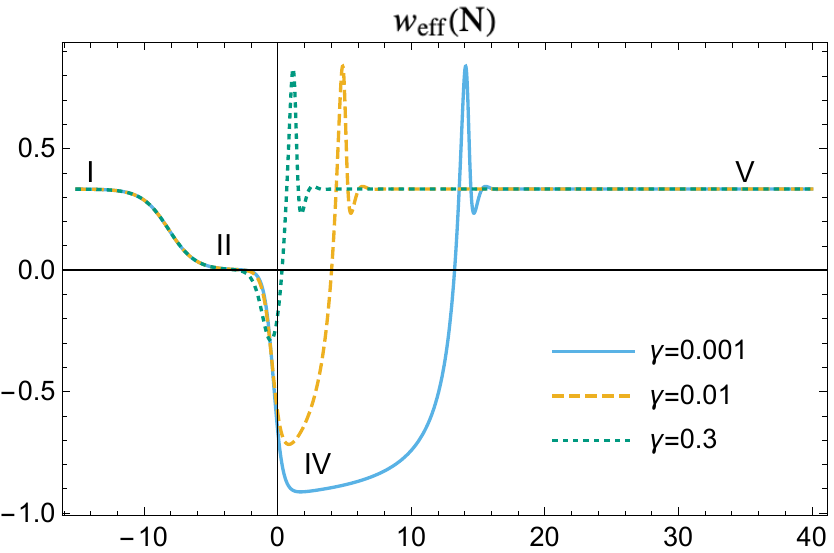}
\end{centering}
\caption{\label{weffnonlocal}Evolution of the effective equation-of-state parameter $w_{\rm eff}$ along path A (left panel) and path B (right panel). The cosmological parameters are fixed to $\Omega_{m}^{0}=0.316$, $H_0=67.24$, and $\Omega_{r}^{0}=0.415\; H_0^{-2}$.
%$\Omega_{r}^{0}=4.15\times10^{-5}h^{-2}$. 
Different curves correspond to different values of $\gamma$; for each value of $\gamma$ and each branch, the initial value $U_0$ is determined by imposing the normalization condition $H(N=0)=H_0$.}
\end{figure}

The system of Eqs.~\eqref{eq:h2ithomega}--\eqref{eq:v1} can then be integrated numerically using the initial conditions specified above for each branch. In Fig.~\ref{weffnonlocal}, we present the evolution of the effective equation-of-state parameter, $w_{\mathrm{eff}}$, along each trajectory. In both cases, saddle points I and II can be clearly identified. They describe the radiation- and matter-dominated regimes, respectively. After the system leaves the vicinity of point II, the nonlocal sector becomes cosmologically relevant, and the differences between the two evolutionary paths become apparent. Remarkably, these paths lead to two unconventional late-time attractors, III and V, whose asymptotic equations of state mimic familiar cosmological regimes without reproducing their underlying physical properties, as we now discuss.

In path A, after passing through the radiation- and matter-dominated saddle points, the system can enter a phantom regime, $w_{\rm eff}<-1$, whose depth depends on the initial condition $U_0$ (or, equivalently, $\gamma$). Different choices of $U_0$ can lead to a stronger or weaker phantom regime, or even to its complete disappearance. After reaching its minimum value, $w_{\rm eff}$ evolves toward $-1$ as the system approaches attractor III. Despite its asymptotic equation of state, this configuration is not a genuine de Sitter state, since the Hubble parameter continues to grow without bound, $H(N)\to\infty$ as $N\to\infty$. We may therefore dub this attractor \textit{``phantom de Sitter''}. At the present epoch ($N=0$), $w_{\rm eff}$ still exhibits a negative slope, indicating that it has not yet reached its minimum.

In path B, the Universe likewise undergoes radiation- and matter-dominated eras as the system passes near saddle points I and II. The evolution then drives the system toward point IV, where it can transiently approach a de Sitter-like regime characterized by $w_{\rm eff}\simeq-1$ and $H(N)\simeq\mathrm{const}$. This point, however, is a saddle rather than an attractor. The system eventually evolves toward attractor V, where $w_{\rm eff}\to1/3$ despite the vanishing radiation density fraction, $\Omega_r\to0$. This peculiar radiation-like expansion, which we may call a \textit{``radiation mirage''}, arises from the nonlocal gravitational dynamics rather than from a genuine radiation component. Indeed, along the physically viable branch satisfying $V_0<1/(3\gamma)-V'_0$, the matter density parameter diverges to $+\infty$, while the effective dark-energy density parameter tends to $-\infty$~\cite{Nersisyan:2016hjh}. The latter does not imply a physical negative-energy matter component, since $\Omega_{\rm DE}$ is an effective quantity associated with the nonlocal gravitational sector.

From a phenomenological perspective, the crucial distinction between the two paths lies in their evolution around the present epoch. Along path B, the minimum of $w_{\rm eff}$ associated with the vicinity of saddle point IV may occur either before or after the present time. Consequently, its present-day slope can be positive or negative, depending on $U_0$ (or, equivalently, $\gamma$), whereas path A exhibits a negative slope for the solutions considered here. Moreover, $w_{\rm eff}$ does not enter a phantom regime along path B, although the effective dark-energy equation of state may still satisfy $w_{\rm DE}<-1$, since $w_{\rm eff}$ and $w_{\rm DE}$ characterize different quantities.

%%%%%%%%%%%%%%%%%%%%%%%%%%%%%%%%%%%%%%%%%%%%%%%%%%%%%%%%%%%%%%%%%%%%%%%
%%%%%%%%%%%%%%%%%%%%%%%%%%%%%%%%%%%%%%%%%%%%%%%%%%%%%%%%%%%%%%%%%%%%%%%

Having analyzed the numerical solutions along both paths, we now compare the phenomenology of the nonlocal model with that of $\Lambda$CDM and the CPL parametrization \cite{Chevallier:2000qy,Linder:2002et}. The equation of state of dark-energy and the associated energy density in $\Lambda$CDM are given by
\begin{equation}
w_{\rm DE}=-1, \qquad \quad  \rho_{\rm DE}=\rho_{\rm DE}^{0},
\end{equation}
while, in the CPL parametrization \cite{Chevallier:2000qy,Linder:2002et}, we have 
\begin{equation}
w_{\rm DE}(a)=w_0+w_a(1-a), \qquad \quad  \rho_{\rm DE}(a)=\rho_{\rm DE}^{0}a^{-3(1+w_0+w_a)}e^{-3w_a(1-a)}.
\end{equation}
Here, $w_0$ represents the present value of the dark-energy equation-of-state parameter at $a=1$, while $w_a$ parametrizes its evolution with the scale factor. Consequently, over the past interval $0<a\leq1$, $w_{\rm DE}(a)$ interpolates between $w_0+w_a$ at early times and $w_0$ at the present cosmological epoch. 

\begin{figure}
\begin{centering}
\includegraphics[height=5.6cm]{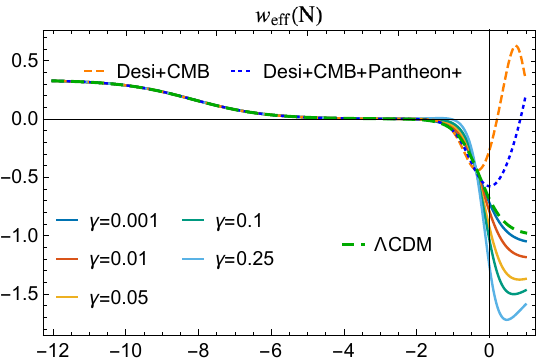}\;\qquad\;
\includegraphics[height=5.6cm]{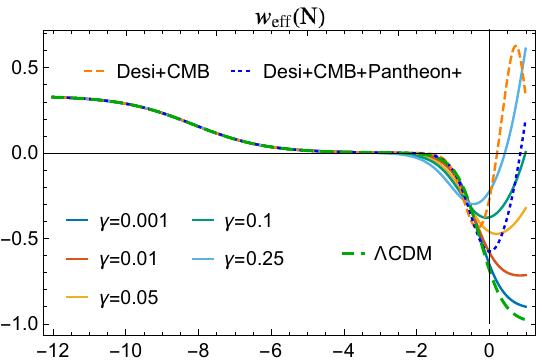}
\end{centering}
\caption{\label{Comparativeweff}Evolution of the effective equation-of-state parameter $w_{\rm eff}$ in $\Lambda$CDM, the CPL parametrization, and the nonlocal model along path A (left panel) and path B (right panel). Dotted curves correspond to CPL, while solid curves represent the nonlocal model. For the nonlocal model and $\Lambda$CDM, we fix $H_0=67.24$, $\Omega_{m}^{0}=0.316$, and $\Omega_{r}^{0}=0.415 H_0^{-2}$. For CPL, we use representative best-fit values of $(\Omega_{m}^{0},H_0,w_0,w_a)$ obtained from the DESI DR2 analysis of Ref.~\cite{DESI:2025zgx}. }
\end{figure}

A direct comparison among the three models can be performed at the level of the effective equation-of-state parameter $w_{\mathrm{eff}}$, defined in Eq.~\eqref{eq:weffdef-1}. Since our goal is to derive observational constraints, we restrict the analysis to the past evolution, which is directly probed by the datasets considered below. In Fig.~\ref{Comparativeweff}, we compare the evolution of $w_{\mathrm{eff}}$ in the three scenarios, showing several values of $\gamma$ for the nonlocal model and representative choices of $(w_0,w_a)$ for CPL.

Path A solutions can produce an evolution of $w_{\mathrm{eff}}$ close to that of $\Lambda$CDM, particularly as $\gamma\rightarrow0$. For the examples displayed in Fig.~\ref{Comparativeweff}, however, the negative present-day slope characteristic of path A differs from the CPL histories considered there, which have already passed their minimum and exhibit a positive slope at $N=0$. This distinction should not be interpreted as a general incompatibility between path A and the entire CPL parameter space, since the sign and magnitude of the CPL slope depend on $(w_0,w_a)$.

Path B is, however, more flexible in the region of interest: for intermediate values of $\gamma$, it can reproduce CPL-like histories, whereas in the limit $\gamma\rightarrow0$ it approaches the $\Lambda$CDM expansion history. Thus, path B provides a continuous family of expansion histories interpolating between behavior close to $\Lambda$CDM and that associated with selected regions of the CPL parameter space. In what follows, we therefore focus on path B, whose effective equation-of-state parameter can vary slowly around the present epoch and whose present-day slope can have either sign. This makes it the branch most relevant for comparison with dynamical-dark-energy parametrizations.

%%%%%%%%%%%%%%%%%%%%%%%%%%%%%%%%%%%%%%%%%%%%%%%%%%%%%%%%%%%%%%%%%%%%%%%
%%%%%%%%%%%%%%%%%%%%%%%%%%%%%%%%%%%%%%%%%%%%%%%%%%%%%%%%%%%%%%%%%%%%%%%

To further quantify the comparison between the nonlocal and the CPL frameworks, we compute the effective CPL parameters, $w_0$ and $w_a$, from the nonlocal theory. For the CPL parametrization, these quantities can be extracted from the present value and first derivative of $w_{\rm eff}$. In particular,
\begin{eqnarray}
w_0=\frac{3\left.w_{\rm eff}\right|_{N=0}-\Omega_{r}^{0}}{3(1-\Omega_{m}^{0}-\Omega_{r}^{0})}\; ,
\label{eqw0}
\end{eqnarray}
\begin{eqnarray}
w_a=2w_0\Omega_{r}^{0}
-3w_0^2(\Omega_{m}^{0}+\Omega_{r}^{0})
+\frac{2\left.\xi'\right|_{N=0}-\Omega_{r}^{0}(1-\Omega_{r}^{0})}
{3(1-\Omega_{m}^{0}-\Omega_{r}^{0})}\;.
\label{eqwa}
\end{eqnarray}
Applying these expressions to the nonlocal model, and fixing $\Omega_m^0$ and $H_0$, each value of $\gamma$ uniquely determines, within a given branch, a pair $(w_0,w_a)$ reproducing the present value and slope of $w_{\rm eff}$. These effective CPL parameters should be understood as local descriptors of the nonlocal expansion history, rather than as independent parameters of the underlying theory. The mapping guarantees agreement between the two descriptions only to first order around the present epoch, without implying an exact correspondence throughout their cosmological evolution. Nevertheless, it provides a useful phenomenological representation of the correlations imposed by the nonlocal dynamics.\footnote{The accuracy of the local CPL mapping can be understood directly from the background evolution. By construction, the nonlocal and CPL descriptions share the same present value and first derivative of $w_{\rm eff}$, implying that their difference starts at second order in an expansion around $N=0$. Since $d\ln H/dN=-3(1+w_{\rm eff})/2$, their logarithmic Hubble rates, normalized to the same present value, differ only at third order in $N$. This explains the close agreement between the expansion histories near the present epoch, while also emphasizing that the local matching does not guarantee accuracy over the full redshift range probed by cosmological observations.}

\begin{figure}
\begin{centering}
\includegraphics[height=5.5cm]{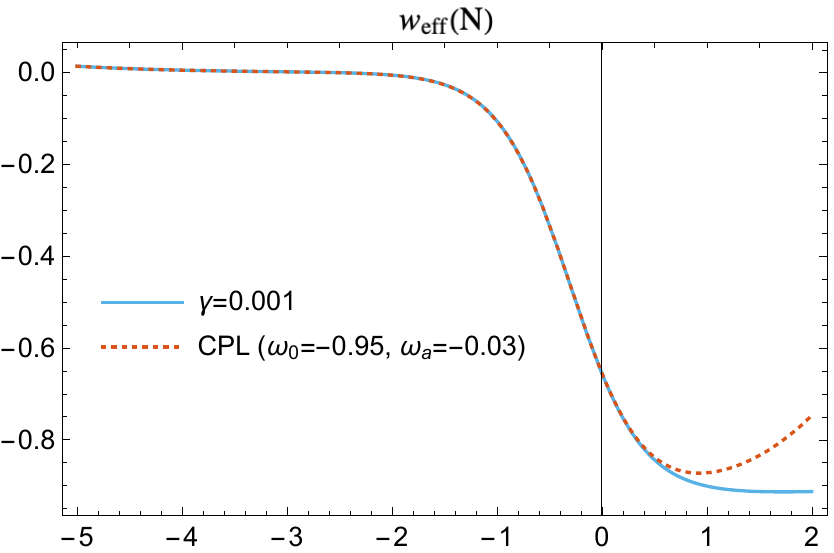}
\includegraphics[height=5.5cm]{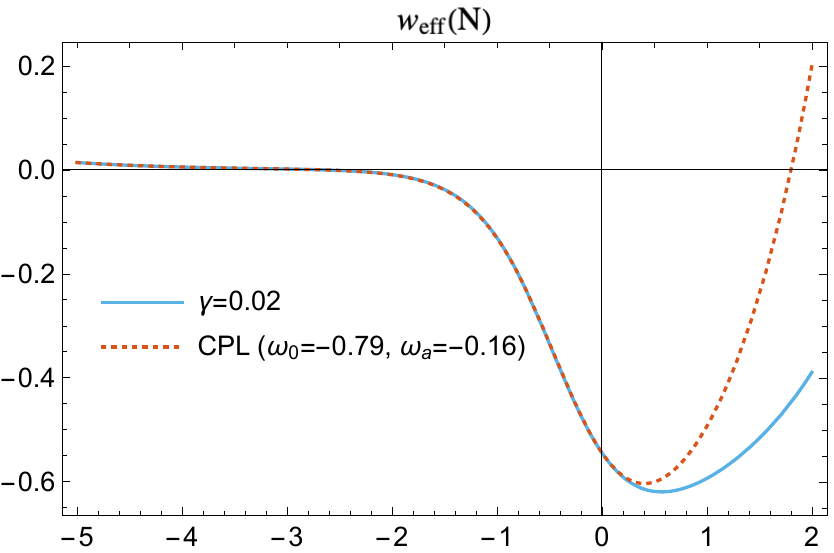}
\caption{Comparison between path B solutions of the nonlocal model and their locally matched CPL descriptions. The left panel shows $w_{\rm eff}$ for the nonlocal solution with $U_0=-70.23$ and for the CPL model with $(w_0,w_a)=(-0.952,-0.033)$. The right panel shows the corresponding comparison for $U_0=-29.70$ and $(w_0,w_a)=(-0.791,-0.156)$. The agreement around the present epoch follows from the construction of the effective CPL parameters, whereas deviations at earlier or later times quantify the limitations of the local mapping.}
\label{graf-Comparison}
\end{centering}
\end{figure}

As an illustrative example, we fix the cosmological parameters and examine the effective CPL parameters associated with two path B solutions, together with the corresponding expansion histories shown in Fig.~\ref{graf-Comparison}. We first consider the free parameters of the nonlocal model: $\Omega_{m}^{0}=0.3114,\; H_0=67.51,\; \gamma=0.001$ , with the associated value $U_0=-70.2278$ selected by the normalization condition $H(N=0)=H_0$ along path B. Evaluating Eqs.~\eqref{eqw0} and \eqref{eqwa}, we obtain the corresponding CPL parameters $w_0=-0.952$ and $w_a=-0.033$. Repeating the procedure for $\gamma=0.02$, which corresponds to $U_0=-29.6991$, while keeping $\Omega_{m}^{0}$ and $H_0$ fixed, yields $w_0=-0.791$ and $w_a=-0.156$. In both cases, $w_a$ is negative and comparatively small in magnitude, particularly for the solution with $\gamma=0.001$.
%%%%%%%%%%%%%%%%%%%%%%%%%%%%%%%%%%%%%%%%%%%%%%%%%%%%%%%%%%%%%%%%%%%%%%%
%%%%%%%%%%%%%%%%%%%%%%%%%%%%%%%%%%%%%%%%%%%%%%%%%%%%%%%%%%%%%%%%%%%%%%%
However, as discussed above, both models undergo the standard radiation- and matter-dominated eras in the past, during which the effects of dark energy and the nonlocal contributions are negligible. As a result, once the values of $(w_0,w_a)$ are chosen to match the present-day behavior of $w_{\mathrm{eff}}$, the two models also exhibit a very similar evolution at earlier times. This agreement is illustrated in Fig.~\ref{graf-Comparison}. It naturally raises the question of whether the MM model can reproduce the behavior favored by DESI for an appropriate choice of its free parameters.

It is important to emphasize, however, that for fixed $\Omega_{m}^{0}$ and $H_0$, the nonlocal model predicts a unique pair $(w_0,w_a)$ for each value of $\gamma$ within a given branch. Hence, $w_0$ and $w_a$ are not independent quantities, but are linked by the underlying dynamics of the theory.  Consequently, the nonlocal model explores only a restricted one-dimensional subset of the full CPL parameter space, in contrast to the CPL parametrization, where $w_0$ and $w_a$ are treated as independent free parameters. Therefore, not every CPL scenario can be realized within the nonlocal framework.

\begin{figure}
\begin{centering}
\includegraphics[height=4.5cm]{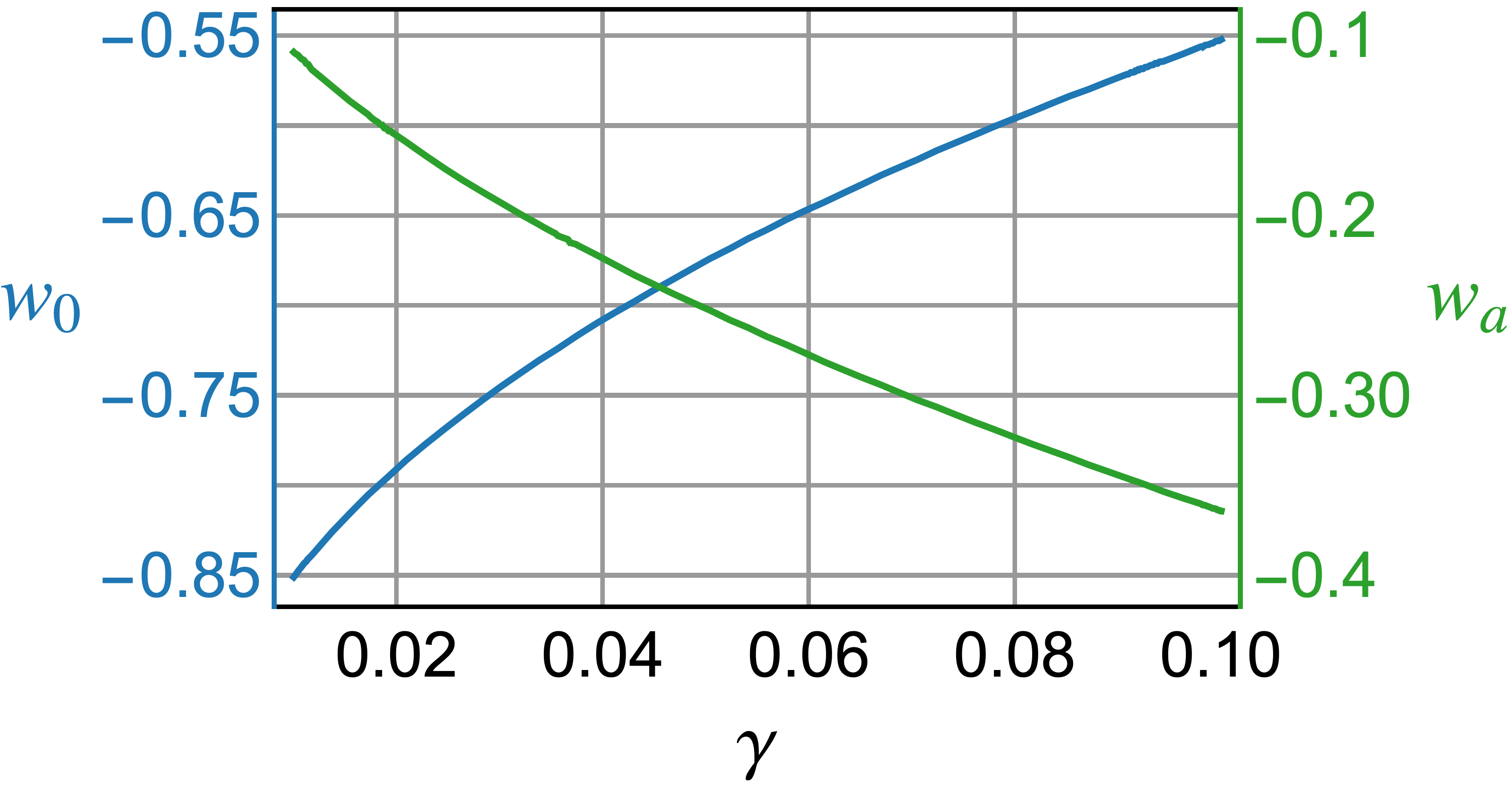}
\includegraphics[height=4.9cm]{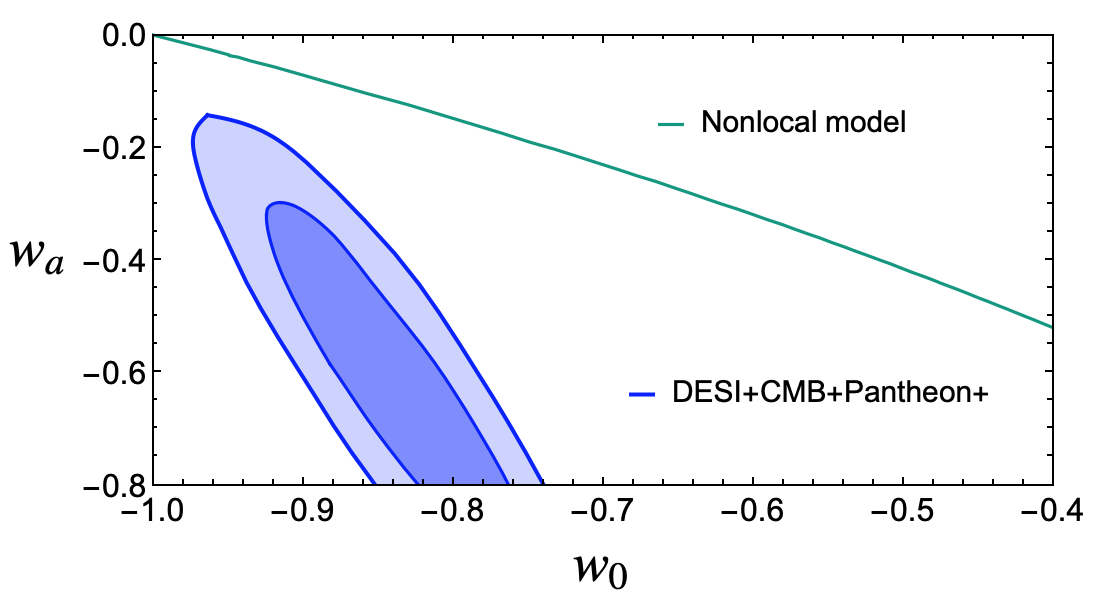}
\end{centering}
\caption{Left: locally matched CPL parameters $w_0$ and $w_a$ predicted by path B as functions of the nonlocal parameter $\gamma$, for fixed $\Omega_m^0=0.3114$ and $H_0=67.51$. Right: corresponding one-dimensional trajectory in the $(w_0,w_a)$ plane, compared with the 68\% and 95\% confidence regions obtained from the DESI + CMB + Pantheon+ for the CPL analysis. The nonlocal model explores only a one-dimensional subset of the full phenomenological parameter space.}
\label{gammaVSw}
\end{figure}

This one-to-one correspondence is illustrated in the left panel of Fig.~\ref{gammaVSw}, which shows the relation between $\gamma$ and the corresponding values of $(w_0,w_a)$ for fixed cosmological parameters $\Omega_{m}^{0}=0.3114$ and $H_0=67.51$. As $\gamma$ increases, $w_0$ increases, whereas $w_a$ decreases. In the right panel of Fig.~\ref{gammaVSw}, we eliminate $\gamma$ and combine both curves to obtain the theoretical relation between $w_0$ and $w_a$ predicted by the nonlocal model. The resulting trajectory is then compared with the results of  DESI constraints \cite{DESI:2025zgx} in the $(w_0,w_a)$ plane obtained within the CPL parametrization.

The shape of this trajectory admits a simple local interpretation close to its $\Lambda$CDM-like endpoint. As derived in Appendix~\ref{app:CPL_mapping}, it can be expanded as
\begin{equation}
w_a=-\alpha_{\rm NL}(1+w_0)+\mathcal{O}\!\left((1+w_0)^2\right),
\label{eq:local_consistency}
\end{equation}
where $\alpha_{\rm NL}>0$ determines the direction along which path B departs from $\Lambda$CDM in the CPL plane. As shown in Appendix~\ref{app:CPL_mapping}, an asymptotic expansion of the background equations in powers of $\sqrt{\gamma}$ yields $\alpha_{\rm NL}\simeq0.686$ for the fiducial cosmological parameters considered here. Equation~\eqref{eq:local_consistency} therefore provides a quantitative consistency relation of the nonlocal model: the effective parameters enter the quadrant $w_0>-1$, $w_a<0$, but cannot vary independently.

The significance of this relation is not simply that the nonlocal model occupies a restricted region of the CPL plane, but that it predicts a specific direction of departure from $\Lambda$CDM. In particular, the evolution toward more negative values of $w_a$ is necessarily accompanied by an increase in $w_0$, preventing the two parameters from adjusting independently to the observationally preferred values. The orientation of the theoretical trajectory thus provides a direct connection between the underlying gravitational dynamics and the model's ability to reproduce the phenomenology suggested by DESI. 

For the fiducial cosmological parameters adopted in Figs.~\ref{graf-Comparison} and \ref{gammaVSw}, the theoretical trajectory does not intersect the region preferred by the unrestricted CPL fit. This comparison is nevertheless illustrative, since the effective CPL parameters are defined through a local matching of the expansion history, whereas the observational contours are inferred from its full redshift dependence. Moreover, variations of the standard cosmological parameters may modify the position and width of the theoretical trajectory. A definitive assessment therefore requires fitting the nonlocal background evolution directly to the cosmological data, as we do in the following section.

%%%%%%%%%%%%%%%%%%%%%%%%%%%%%%%%%%%%%%%%%%%%%%%%%%%%%%%%%%%%%%%%%%%%%%%
%%%%%%%%%%%%%%%%%%%%%%%%%%%%%%%%%%%%%%%%%%%%%%%%%%%%%%%%%%%%%%%%%%%%%%%

%%%%%%%%%%%%%%%%%%%%%%%%%%%%%%%%%%%%%%%%%%%%%%%%%%%%%
\section{Constraints from DESI and complementary datasets}
\label{sec:constraints}
%%%%%%%%%%%%%%%%%%%%%%%%%%%%%%%%%%%%%%%%%%%%%%%%%%%%%

We fit the path-B model directly to the cosmological likelihoods, sampling the nonlocal parameter $\gamma$ together with the standard cosmological parameters. We recall that the initial value $U_0$ is not sampled independently but is determined at each point by imposing the normalization condition $H(N=0)=H_0$, while retaining the solution corresponding to path B.

We implemented the background cosmological dynamics of path B in the Boltzmann code CLASS \cite{Lesgourgues:2011re}  and interfaced it with the Markov chain Monte Carlo sampler \texttt{MontePython} \cite{brinckmann2018montepython}. While the nonlocal background evolution is consistently incorporated, cosmological perturbations are treated within an effective smooth-dark-energy approximation. This distinction is particularly relevant because BAO and supernova distances primarily constrain the homogeneous expansion history, whereas CMB temperature and polarization observables also depend on perturbation dynamics. Nevertheless, since the observationally favored solutions exhibit a background evolution close to that of $\Lambda$CDM over the redshift range probed by current data, one may reasonably expect the corrections arising from a fully consistent treatment of nonlocal perturbations to remain relatively modest in this region of parameter space, although this remains to be verified explicitly. The constraints reported below should therefore be interpreted as conditional on the adopted effective perturbation prescription. A more complete assessment would require implementing the modified perturbation equations of the theory.

For every point in parameter space, the auxiliary-field equations are integrated from $N_{\rm ini}=-20.7233$, imposing
\begin{equation}
V(N_{\rm ini})=0\,,\qquad V'(N_{\rm ini})=0\,,\qquad U'(N_{\rm ini})=0\,,
\label{eq:MCMC_initial_conditions}
\end{equation}
while $U(N_{\rm ini})\equiv U_0$ is determined numerically by requiring the normalization condition. Among the two solutions of the shooting procedure, we retain the one belonging to path B, namely $U_0<\bar U(\gamma)$. We then sample the six standard cosmological parameters
\begin{equation}
\left\{\omega_b,\omega_{cdm},\theta_s,\tau_{\rm reio},\ln(10^{10}A_s),n_s\right\}\,,
\label{eq:standard_parameter_set}
\end{equation}
together with the nonlocal parameter $\gamma$. Here $\omega_b\equiv\Omega^0_b(H_0/100)^2$ and $\omega_{cdm}\equiv\Omega^0_{cdm}(H_0/100)^2$ are the physical baryon and cold-dark-matter densities, respectively, while $\theta_s$ denotes the acoustic-scale parameter used in the chains. The primordial scalar spectrum is taken to be a power law with amplitude $A_s$ and spectral tilt $n_s$. 

In Table~\ref{tab:priors}, we summarize the cosmological parameters fitted in the different runs, together with the priors imposed on them. It should be noted that these priors are  constrained by the grid over which the normalization of $H_0$ and $U_0$ was performed, which depends on the parameter space of $H_0$, $\Omega^0_m$ and $\gamma$. These values were chosen close to those expected for a broad range of cosmological models and within a relatively narrow region, allowing us to construct a finer grid to accurately capture the dependence on $\gamma$. The priors for the remaining cosmological parameters are assumed to be flat and unbounded, i.e., flat priors without explicitly imposed finite bounds. In addition, we impose $\gamma\geq0$ and restrict its range to values for which the path-B shooting procedure yields a regular cosmological evolution over the entire redshift interval relevant to the likelihood analysis. An upper prior boundary $\gamma<0.2$ is adopted to restrict the exploration to models whose background evolution is \textit{a priori} close to both $\Lambda$CDM and the DESI CPL data fit. Since the $\Lambda$CDM-like expansion history is approached through the singular scaling limit $\gamma\rightarrow0$ and $U_0\rightarrow-\infty$, the point $\gamma=0$ is not integrated directly. Instead, it is understood as the limiting solution approached continuously as $\gamma$ decreases.

%%%%%%%%%%%%%%%%%%%%%%%%%%%%%%%%%%%%%%%%%%%%%%%%%%%%%%%%%%%%%%%%%%%%%%%%%%%%%%%%%%%%%%%%%%%%%
\begin{table}
\centering
\begin{tabular}{lcc}
\hline\hline
Parameter & Prior & Description \\
\hline
$\Omega^0_m$ & $\mathcal{U}[0.27,0.35]$ & Matter density \\
$H_0$ & $\mathcal{U}[60,80]$ & Hubble parameter ($\text{km}\; \text{s}^{-1} \;\text{Mpc}^{-1}$)  \\
\hline
$\omega_b$ & Flat (Unbounded) & Baryon density \\
$\omega_{cdm}$ &  Flat (Unbounded) & Cold-dark-matter density \\
$100 \theta_s$ &  Flat (Unbounded) & Angular acoustic scale \\
$\tau_{\rm reio}$ & Flat (Unbounded) & Reionization optical depth \\
$\ln(10^{10}A_s)$ &  Flat (Unbounded) & Scalar amplitude \\
$n_s$ &  Flat (Unbounded) & Scalar spectral index \\
$\gamma$ & $\mathcal{U}[0.001,0.2]$ & Nonlocal parameter \\
\hline\hline
\end{tabular}
\caption{Cosmological parameters and priors used in the MCMC analysis. The nuisance parameters associated with the different likelihoods are varied according to the prescriptions of the corresponding collaborations. Unbounded denotes a flat prior without explicitly imposed finite bounds.
}
\label{tab:priors}
\end{table}
%%%%%%%%%%%%%%%%%%%%%%%%%%%%%%%%%%%%%%%%%%%%%%%%%%%%%%%%%%%%%%%%%%%%%%%%%%%%%%%%%%%%%%%%%%%%%

%%%%%%%%%%%%%%%%%%%%%%%%%%%%%%%%%%%%%%%%%%%%%%%%%%%%%
\subsection{Datasets and statistical methodology} \label{subsec:datasets}
%%%%%%%%%%%%%%%%%%%%%%%%%%%%%%%%%%%%%%%%%%%%%%%%%%%%%

To constrain the model parameters, we perform a comprehensive MCMC analysis based on the following cosmological datasets:
%%%%%%%%%%%%%%%%%%%%%%%%%%%%%%%%%%%%%%%%%%%%%%%%%%%%%%%%%%%%%%%%%%%%%%%
\begin{itemize}
\item \textbf{Planck 2018:} The CMB likelihood combines the Planck TT, EE, and TE power spectra, using \texttt{simall} and \texttt{Commander} for $l<30$ and \texttt{Plik} for $l\geq30$ \cite{Planck:2018vyg}.
\item \textbf{DESI DR2 BAO:} We consider the Baryon Acoustic Oscillation (BAO) measurements from the second DESI data release \cite{DESI:2025zgx}. The dataset combines information from several tracers, including galaxies, quasars, and the Lyman-$\alpha$ forest, and provides both isotropic and anisotropic constraints on cosmological distances across the redshift range $0.295 \leq z \leq 2.330$, divided into nine redshift bins.
\item \textbf{Pantheon+:} We use the Pantheon+ compilation of 1550 unique Type Ia supernovae, covering the redshift range $0.001 \leq z \leq 2.26$.
\end{itemize}
%%%%%%%%%%%%%%%%%%%%%%%%%%%%%%%%%%%%%%%%%%%%%%%%%%%%%%%%%%%%%%%%%%%%%%%

We denote this combination by Planck+DESI\ DR2$+$Pantheon+. For comparison, we also analyze DESI+CMB and CMB alone. These additional combinations allow us to assess the dependence of the inferred constraints on the adopted supernova compilation.

We require the Gelman--Rubin convergence criterion to satisfy $R-1<0.01$ for all sampled parameters \cite{Gelman:1992zz}. The quoted intervals correspond to marginalized 68\% and 95\% credible regions unless stated otherwise. Best-fit points and minimum effective chi-squared values are obtained using the Sequential Least Squares Programming (SLSQP) optimizer implemented within \texttt{MontePython.}

%%%%%%%%%%%%%%%%%%%%%%%%%%%%%%%%%%%%%%%%%%%%%%%%%%%%%
\subsection{Constraints on the nonlocal model} \label{subsec:nonlocal_constraints}
%%%%%%%%%%%%%%%%%%%%%%%%%%%%%%%%%%%%%%%%%%%%%%%%%%%%%
%%%%%%%%%%%%%%%%%%%%%%%%%%%%%%%%%%%%%%%%%%%%%%%%%%%%%%%%%%%%%%%%%%%%%%%
\begin{figure}
    \centering
    \includegraphics[width=0.9\linewidth]{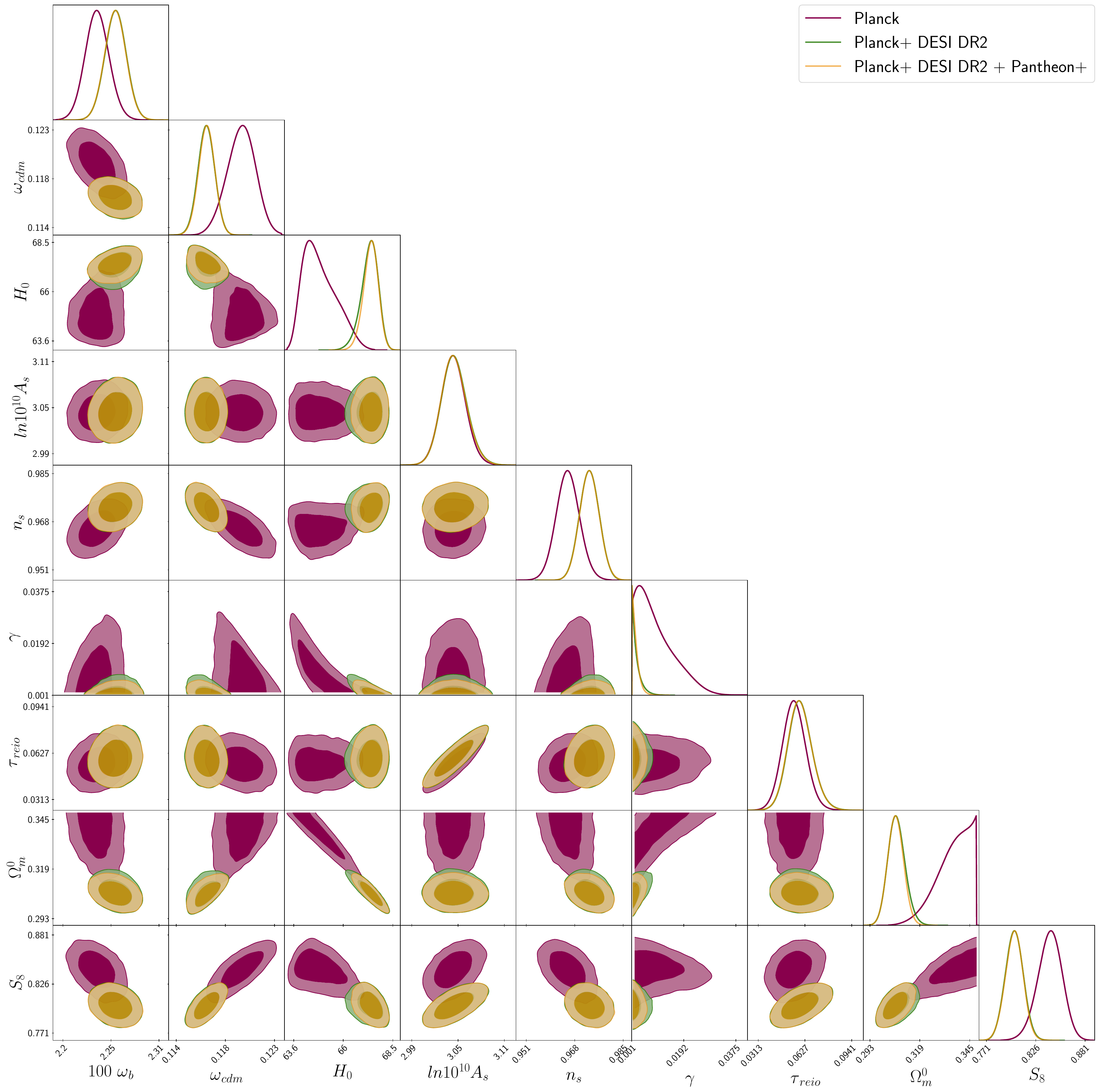}
    \caption{Marginalized posterior distributions for path B of the nonlocal model obtained from the combined Planck, DESI DR2 BAO, and Pantheon+ Type Ia supernova datasets.}
    \label{fig:MCMC_path_B}
\end{figure}
%%%%%%%%%%%%%%%%%%%%%%%%%%%%%%%%%%%%%%%%%%%%%%%%%%%%%%%%%%%%%%%%%%%%%%%

The marginalized posterior distributions for the baseline dataset are shown in Fig.~\ref{fig:MCMC_path_B} and Table~\ref{MCMC_table}. The standard cosmological parameters remain broadly consistent with their values inferred in $\Lambda$CDM. The principal effect of the nonlocal extension is encoded in the posterior of $\gamma$, which is peaked toward the lower prior boundary. For the Planck+DESI\ DR2+Pantheon+ dataset combination, we obtain 
\begin{equation}
\gamma<0.00495\qquad (95\%~{\rm C.L.})\,.
\label{eq:gamma_constraint}
\end{equation}
The fact that the posterior accumulates close to $\gamma=0$ indicates that the combined data favor the region in which the background expansion approaches its $\Lambda$CDM-like limit. This behavior can be understood from the restricted relation between the effective CPL parameters predicted by the model. Increasing $\gamma$ moves the path-B solutions away from the $\Lambda$CDM region, increasing $w_0$ while making $w_a$ more negative, as shown in Fig.~\ref{gammaVSw}. The resulting trajectory does not follow the principal degeneracy direction of the phenomenological CPL constraints. Consequently, the values of $\gamma$ that generate an appreciable departure from $w_{\rm DE}=-1$ cannot simultaneously reproduce the combination of $w_0$ and $w_a$ preferred by an unrestricted $w_0w_a$CDM fit.

The resulting constraints suggest that variations of the standard cosmological parameters do not provide sufficient freedom to reconcile the path-B background evolution with the expansion histories preferred by the unrestricted CPL fit. The local consistency relation~\eqref{eq:local_consistency} provides a dynamical interpretation of this result: moving toward more negative values of $w_a$ necessarily drives $w_0$ farther above $-1$, eventually degrading the fit to the combined data. Consequently, the nonlocal parameter is only weakly correlated with the other parameters, except for the Planck data set where we find a slight anticorrelation with $H_0$, indicating that larger values of $\gamma$ are compatible with smaller values of $H_0$, and a stronger correlation with $\Omega^0_m$. However, when all data sets are combined, these correlations become negligible, and the nonlocal model instead exhibits a tendency towards the $\Lambda$CDM limit. In fact, the inferred Hubble constant, $H_0=67.39^{+0.399}_{-0.335}\,\,{\rm km\,s^{-1}\,Mpc^{-1}}$ is slightly shifted downward relative to the corresponding $\Lambda$CDM result. Nevertheless, this shift is not sufficient to activate the new dynamical features of the nonlocal model required to accommodate the data.

On the other hand, the sharp edge in the Planck-only $\Omega_m^0$ posterior in Fig.~\ref{fig:MCMC_path_B} is caused by the upper boundary of the grid used for the calculation of $U_0$, which translates into the upper prior boundary $\Omega_m^0=0.35$. The combined-dataset posteriors remain safely within the interpolation range.

For fixed standard cosmological parameters, the mapping between $\gamma$ and $(w_0,w_a)$ defines a one-dimensional trajectory in the CPL plane. Allowing these parameters to vary broadens this trajectory into a family of closely related curves, whose observationally relevant region remains constrained by the correlations imposed by the nonlocal dynamics.

%%%%%%%%%%%%%%%%%%%%%%%%%%%%%%%%%%%%%%%%%%%%%%%%%%%%%%%%%%%%%%%%%%%%%%%%%%%%%%%%%%%%%%
\begin{table}
\centering
\begin{tabular}{c|cc|cc|cc}
\toprule
 & \multicolumn{2}{c|}{Planck}  & \multicolumn{2}{c|}{Planck + DESI DR2 BAO } & \multicolumn{2}{c}{\makecell{Planck + DESI DR2 BAO +  \\  Pantheon+}}    \\
 \midrule
Parameter & $\Lambda$CDM & Nonlocal& $\Lambda$CDM & Nonlocal& $\Lambda$CDM & Nonlocal  \\
\midrule
$100\omega_b$
 & $2.240\pm0.015$
  & $2.238\pm {0.015}$ & $2.250\pm {0.013}$
  & $2.260\pm {0.013}$ & $2.250\pm {0.013}$
  & $2.260\pm {0.013}$ \\
$\omega_{\rm cdm}$
  & $0.1200\pm0.0014$
  & $0.1198\pm {0.0013}$ & $0.1180^{+0.00068}_{-0.00069}$
  & $0.11668^{+0.00075}_{-0.00073}$ & $0.118\pm {0.001}$
  & $0.11670^{+0.00074}_{-0.00073}$ \\
$100\theta_s$
  & $1.04\pm0.00030$
  & $1.04191\pm {0.00030}$ & $1.0423\pm {0.00028}$
  & $1.0422^{+0.00030}_{-0.00028}$& $1.04\pm0.000278$
  & $1.04219\pm {+0.00028}$ \\
$\ln(10^{10}A_s)$
   & $3.05^{+0.016}_{-0.017}$
  & $3.046^{+0.015}_{-0.017}$ & $3.050^{+0.016}_{-0.018}$
  & $3.048^{+0.016}_{-0.018}$ & $3.05^{+0.016}_{-0.018}$
  & $3.047^{+0.0162}_{-0.0180}$ \\
$n_s$
   & $0.965^{+0.0045}_{-0.0046}$
  & $0.9657^{+0.0043}_{-0.0044}$ & $0.9710^{+0.0035}_{-0.0034}$
  & $0.9733^{+0.0035}_{-0.0035}$ & $0.971\pm0.0034$
  & $0.9733^{+0.0035}_{-0.0036}$ \\
$\tau_{\rm reio}$
  & $0.0545^{+0.0076}_{-0.0082}$
  & $0.0554^{+0.0074}_{-0.0084}$ & $0.0579^{+0.0076}_{-0.0085}$
  & $0.0596^{+0.0077}_{-0.0088}$  & $0.0575^{+0.0076}_{-0.0085}$
   & $0.0594^{+0.0076}_{-0.0088}$ \\
$\gamma$ ($95\%$ C.L.)
  & $-$
  & $<0.0222$  & $-$
  & $<0.00624$ & $-$
   & $<0.00495$ \\
\midrule
$H_0$
  & $67.3^{+0.63}_{-0.64}$
  & $64.91^{+0.59}_{-1.05}$ & $68.50^{+0.31}_{-0.30}$
  & $67.33^{+0.49}_{-0.36}$ & $68.4\pm0.302$
  & $67.39^{+0.40}_{-0.34}$ \\
$\sigma_8$
  & $0.811^{+0.0077}_{-0.0076}$
  & $0.7937\pm {+0.010}$ & $0.8040^{+0.0069}_{-0.0075}$
  & $0.7926^{+0.0082}_{-0.0081}$ & $0.805^{+0.0069}_{-0.0075}$
  & $0.7930^{+0.0077}_{-0.0079}$ \\
\midrule
$\chi^2_{\min}$ & $2772.09$ & $2771.41$ & $2788.677$ & $2786.129$ & $4200.94$ & $4198.286$ \\
$\Delta\chi^2_{\min}$ & $0$ & $-0.68$ & $0$ & $-2.548$ & $0$ & $-2.654$ \\
\bottomrule
\end{tabular}
\caption{Constraints on the cosmological parameters of $\Lambda$CDM and the path-B nonlocal model with the best-fit effective chi-squared contributions. Both models are evaluated using the same likelihood combination and nuisance-parameter treatment.}
\label{MCMC_table}
\end{table}
%%%%%%%%%%%%%%%%%%%%%%%%%%%%%%%%%%%%%%%%%%%%%%%%%%%%%%%%%%%%%%%%%%%%%%%%%%%%%%%%%%%%%%

%%%%%%%%%%%%%%%%%%%%%%%%%%%%%%%%%%%%%%%%%%%%%%%%%%%%%
\subsection{Goodness of fit and comparison with $\Lambda$CDM} \label{subsec:model_comparison}
%%%%%%%%%%%%%%%%%%%%%%%%%%%%%%%%%%%%%%%%%%%%%%%%%%%%%

To assess whether the additional nonlocal degree of freedom improves the description of the data, we compare the minimum effective chi-squared with that obtained in $\Lambda$CDM using the same likelihood combination. From a combined Planck, DESI DR2 and Pantheon+ dataset, we find 
\begin{equation}
\Delta\chi^2_{\rm min}\equiv\chi^2_{{\rm min},{\rm nonlocal}}-\chi^2_{{\rm min},\Lambda{\rm CDM}}=-2.654\,.
\label{eq:delta_chi2}
\end{equation}

The minimum effective $\chi$-squared values for the different dataset combinations are reported in Table~\ref{MCMC_table}. Since the nonlocal model contains one additional sampled parameter relative to $\Lambda$CDM, a small improvement of the best-fit chi-squared would not by itself constitute evidence in its favor. For reference, the difference in the Akaike information criterion is
\begin{equation}
\Delta{\rm AIC}=\Delta\chi^2_{\rm min}+2=-0.654\,.
\label{eq:delta_AIC}
\end{equation}

For the Planck+DESI DR2+Pantheon+ dataset combination, the negative value of $\Delta\mathrm{AIC}$ indicates a marginal preference for the nonlocal model, even after accounting for the additional parameter. However, its small magnitude implies that this preference is not statistically meaningful. This result is also consistent with the behavior of the nonlocal parameter found in the observational analysis. The constraints tend to push $\gamma$ towards small values, precisely the regime in which the nonlocal model approaches $\Lambda$CDM. Consequently, the modest improvement in $\chi^2_{\rm min}$ does not appear to be associated with a preference for a genuinely nonlocal cosmological regime. Instead, the additional degree of freedom remains weakly activated by the data, with the best-fit region lying close to the $\Lambda$CDM limit. In this sense, the AIC comparison and the posterior constraints provide a consistent picture: the present data do not require the additional nonlocal degree of freedom to improve the fit, while remaining compatible with the nonlocal extension.

%%%%%%%%%%%%%%%%%%%%%%%%%%%%%%%%%%%%%%%%%%%%%%%%%%%%%%%%%%%%%%%%%%%%%%%
\section{Summary and conclusions}\label{concl}
%%%%%%%%%%%%%%%%%%%%%%%%%%%%%%%%%%%%%%%%%%%%%%%%%%%%%%%%%%%%%%%%%%%%

In this work, we have investigated the cosmological dynamics and observational viability of the nonlocal $R\Box^{-2}R$ model, focusing on the class of solutions known as path B. These solutions reproduce the standard radiation- and matter-dominated epochs before approaching a quasi-de Sitter saddle point, where cosmic acceleration can persist for an extended period. Unlike the asymptotic de Sitter-like behavior of path A, the accelerated stage along path B is transient, eventually giving way to a radiation-like regime without radiation. Around the present epoch, however, path B can generate expansion histories resembling those of dynamical dark-energy models.

To characterize these histories, we have constructed a local mapping between the nonlocal background evolution and the effective CPL parameters $(w_0,w_a)$. For fixed standard cosmological parameters, the dynamics restrict these quantities to a one-dimensional trajectory rather than allowing them to vary independently. Close to the $\Lambda$CDM-like limit, we have shown that this trajectory obeys the consistency relation $w_a=-\alpha_{\rm NL}(1+w_0)+\mathcal{O}((1+w_0)^2)$. By expanding the nonlocal background equations in powers of $\sqrt{\gamma}$, we have determined its asymptotic slope, obtaining $\alpha_{\rm NL}\simeq0.686$ for our fiducial cosmological parameters. This result provides a quantitative connection between the underlying gravitational dynamics and the direction of departure from $\Lambda$CDM in the CPL plane.

Although the path-B trajectory enters the quadrant $w_0>-1$, $w_a<0$, also favored by phenomenological CPL fits to DESI, its orientation differs from that preferred by the data. In particular, increasingly negative values of $w_a$ necessarily correspond to values of $w_0$ farther above $-1$, preventing the model from exploring the full range of independently varying CPL parameters. Our likelihood analysis, combining CMB, DESI DR2 BAO, and Pantheon+ observations, confirms that this restricted evolution does not provide sufficient freedom to reproduce the expansion histories preferred by an unrestricted CPL fit. Instead, the posterior favors small values of $\gamma$, driving the allowed solutions toward their $\Lambda$CDM-like limit.

The comparison with $\Lambda$CDM for a combined dataset of Planck+DESI DR2+Pantheon+ yields a modest improvement in the minimum effective chi-squared, $\Delta\chi^2_{\rm min}=-2.654$, corresponding to $\Delta{\rm AIC}=-0.654$ . The latter difference is too small to constitute meaningful statistical evidence in favor of the nonlocal model. These findings should nevertheless be interpreted within the scope of our implementation, which modifies the homogeneous cosmological background while retaining an effective treatment of perturbations. The resulting constraints are therefore conditional on this approximation and on the adopted priors. A complete assessment of the underlying modified-gravity theory would require a consistent implementation of its perturbation equations.

Our analysis is also restricted to the path-B realization defined by the auxiliary-field initial conditions adopted here. Different admissible prescriptions for the nonlocal operator may lead to distinct cosmological histories, and their observational viability remains to be investigated. Extending the present framework to the complete perturbation sector and examining complementary cosmological probes would provide further tests of the model.

More generally, our results illustrate the importance of confronting theoretically motivated models of cosmic acceleration with observations beyond their ability to generate a time-dependent dark-energy equation of state. The relevant question is whether the specific correlations imposed by their underlying dynamics agree with those inferred from cosmological data. In the path-B realization studied here, these correlations favor an observationally allowed region close to $\Lambda$CDM, with no significant evidence for genuinely nonlocal deviations.

\acknowledgments  
This article is based on work from COST Action CA21136 ``Addressing observational tensions in cosmology with systematics and fundamental physics'' (CosmoVerse) supported by COST (European Cooperation in Science and Technology). IA is supported by the Basque Government Grant No.~IT1628-22, Grant No.~PID2021-123226NB-I00 (funded by MCIN/AEI/10.13039/501100011033, by ERDF ``A way of making Europe''), and the Grant Juan de la Cierva funded by MICIU/AEI/10.13039/501100011033 and by ESF+. JR  (ORCID 0000-0001-7545-1533) is supported by a Ram\'on y Cajal contract from the Spanish Ministry of Science and Innovation with Ref. RYC2020-028870-I. This work was supported by the project PID2022-139841NB-I00 of MICIU/AEI/10.13039/501100011033 and FEDER, UE. LA, AC, and SD acknowledge the support by DST-DAAD Indo-German joint research collaboration grant DST/INT/DAAD/P-07/2023(G). AC and SD thank Heidelberg University for its kind hospitality and stimulating research environment, where part of the research work was carried out during the scientific visit.
LA acknowledges support by DFG under Germany's Excellence Strategy EXC 2181/1 - 390900948 (the Heidelberg STRUCTURES Excellence Cluster), under Project 554679582 ``GeoGrav: Cosmological Geometry and Gravity with nonlinear physics'', and  by Tamkeen under the NYUAD Research Institute grant CG014.

%%%%%%%%%%%%%%%%%%%%%%%%%%%%%%%%%%%%%%%%%%%%%%%%%%%%%%%%%%%%%%%%%%%%%%%
\appendix

%%%%%%%%%%%%%%%%%%%%%%%%%%%%%%%%%%%%%%%%%%%%%%%%%%%%%%%%%%%%%%%%%%%%%%%
\section{Local CPL mapping and asymptotic consistency relation}
\label{app:CPL_mapping}
%%%%%%%%%%%%%%%%%%%%%%%%%%%%%%%%%%%%%%%%%%%%%%%%%%%%%

In this appendix, we derive the local mapping between the background evolution of the nonlocal model and the effective Chevallier--Polarski--Linder parameters $(w_0,w_a)$. We then determine the asymptotic behavior of the path-B solutions around their $\Lambda$CDM-like limit and derive the corresponding consistency relation in the CPL plane. In particular, we show that the leading deviations from $\Lambda$CDM can be obtained from a system of linear differential equations, allowing us to determine the tangent of the theoretical trajectory without integrating the full nonlinear system at finite $\gamma$.

%%%%%%%%%%%%%%%%%%%%%%%%%%%%%%%%%%%%%%%%%%%%%%%%%%%%%
\subsection{Exact local mapping}
\label{app:CPL_exact}
%%%%%%%%%%%%%%%%%%%%%%%%%%%%%%%%%%%%%%%%%%%%%%%%%%%%%

For a spatially flat cosmology containing pressureless matter, radiation, and an effective dark-energy component, the total effective equation-of-state parameter can be written as
\begin{equation}
w_{\rm eff}(N)=\frac{\Omega_r(N)}{3} +w_{\rm DE}(N)\Omega_{\rm DE}(N)\,,
\qquad
\Omega_{\rm DE}(N)=1-\Omega_m(N)-\Omega_r(N)\,.
\label{eq:app_weff_decomposition}
\end{equation}
For the CPL parametrization,
\begin{equation}
w_{\rm DE}(a)=w_0+w_a(1-a) =w_0+w_a(1-e^N)\,,
\label{eq:app_CPL_definition}
\end{equation}
and therefore $w_{\rm DE}(0)=w_0$ and $w_{\rm DE}'(0)=-w_a$. Evaluating Eq.~\eqref{eq:app_weff_decomposition} at the present epoch gives
\begin{equation}
w_{{\rm eff},0}=\frac{\Omega_r^0}{3} +w_0\Omega_{\rm DE}^0\,,
\qquad\qquad
\Omega_{\rm DE}^0=1-\Omega_m^0-\Omega_r^0\,,
\label{eq:app_weff_today}
\end{equation}
from which
\begin{equation}
w_0=\frac{3w_{{\rm eff},0}-\Omega_r^0}
{3\left(1-\Omega_m^0-\Omega_r^0\right)}\,.
\label{eq:app_w0_exact}
\end{equation}
Using $w_{\rm eff}=-1-2\xi/3$, with $\xi\equiv H'/H$, this expression may equivalently be written as
\begin{equation}
w_0=\frac{-3-2\xi_0-\Omega_r^0}
{3\left(1-\Omega_m^0-\Omega_r^0\right)}\,.
\label{eq:app_w0_xi}
\end{equation}
To determine $w_a$, we differentiate Eq.~\eqref{eq:app_weff_decomposition},
\begin{equation}
w_{\rm eff}'=\frac{\Omega_r'}{3} +w_{\rm DE}'\Omega_{\rm DE} +w_{\rm DE}\Omega_{\rm DE}'\,.
\label{eq:app_weff_derivative}
\end{equation}
The matter and radiation fractions satisfy
\begin{equation}
\Omega_m'=-(3+2\xi)\Omega_m\,,
\qquad
\Omega_r'=-(4+2\xi)\Omega_r\,,
\label{eq:app_density_evolution}
\end{equation}
while spatial flatness implies
\begin{equation}
\Omega_{\rm DE}'=(3+2\xi)\Omega_m +(4+2\xi)\Omega_r\,.
\label{eq:app_OmegaDE_derivative}
\end{equation}
Evaluating Eq.~\eqref{eq:app_weff_derivative} at $N=0$, using $w_{\rm eff}'=-2\xi'/3$, and solving for $w_a$, we obtain
\begin{equation}
w_a=\frac{\Omega_r^{\prime 0}/3 +w_0\Omega_{\rm DE}^{\prime 0} +2\xi_0'/3}{\Omega_{\rm DE}^0}\,.
\label{eq:app_wa_unsimplified}
\end{equation}
Substituting Eqs.~\eqref{eq:app_density_evolution} and \eqref{eq:app_OmegaDE_derivative}, and eliminating $\xi_0$ using Eq.~\eqref{eq:app_w0_xi}, yields
\begin{equation}
w_a=2w_0\Omega_r^0 -3w_0^2\left(\Omega_m^0+\Omega_r^0\right) +\frac{2\xi_0'-\Omega_r^0(1-\Omega_r^0)}
{3\left(1-\Omega_m^0-\Omega_r^0\right)}\,.
\label{eq:app_wa_exact}
\end{equation}
Equations~\eqref{eq:app_w0_exact} and \eqref{eq:app_wa_exact} show that the effective CPL parameters are determined by the present background evolution and its derivatives, together with the matter and radiation fractions. The mapping is local: it matches the present value and first derivative of the effective dark-energy equation of state, without implying an exact correspondence between the nonlocal and CPL expansion histories at arbitrary redshift. As a consistency check, for a $\Lambda$CDM background one has $w_0=-1$ and $w_a=0$. Neglecting radiation for simplicity, $\xi=-3\Omega_m/2$ and $\xi'=9\Omega_m\Omega_{\rm DE}/2$, so that Eq.~\eqref{eq:app_wa_exact} gives
\begin{equation}
w_a=-3\Omega_m^0 +\frac{2}{3\Omega_{\rm DE}^0}\frac{9}{2}\Omega_m^0\Omega_{\rm DE}^0=0\,.
\label{eq:app_LCDM_check}
\end{equation}

%%%%%%%%%%%%%%%%%%%%%%%%%%%%%%%%%%%%%%%%%%%%%%%%%%%%%
\subsection{Asymptotic expansion of the nonlocal background}
\label{app:CPL_asymptotic_background}
%%%%%%%%%%%%%%%%%%%%%%%%%%%%%%%%%%%%%%%%%%%%%%%%%%%%%

For fixed values of the standard cosmological parameters, each path-B solution determines a unique pair $(w_0,w_a)$. The resulting trajectory is therefore one-dimensional, although its parametrization in terms of $\gamma$ becomes singular at the $\Lambda$CDM-like endpoint. To characterize its local behavior, we introduce $\epsilon\equiv\sqrt{\gamma}$ and expand the background equations in the limit $\epsilon\rightarrow0$. At leading order, the expansion history approaches
\begin{equation}
E^2(N)=\Omega_m^0e^{-3N}
+\Omega_r^0e^{-4N}+\Omega_\Lambda\,,
\qquad \Omega_\Lambda\equiv1-\Omega_m^0-\Omega_r^0\,,
\label{eq:app_LCDM_background}
\end{equation}
where $E(N)$ denotes the normalized Hubble rate in $\Lambda$CDM.

Motivated by the singular behavior of the auxiliary-field initial condition, we consider the asymptotic expansions
\begin{equation}
U(N)=-\frac{C}{\epsilon}+b+u(N)+\mathcal O(\epsilon)\,,
\qquad
V(N)=-\frac{C}{\epsilon}q(N)
+\mathcal O(1)\,,
\label{eq:app_auxiliary_expansion}
\end{equation}
together with
\begin{equation}
h^2(N)=E^2(N)+\epsilon d(N)+\mathcal O(\epsilon^2)\,.
\label{eq:app_Hubble_expansion}
\end{equation}
Here $C$ and $b$ are constants, while $u(N)$ and $q(N)$ describe the leading time-dependent contributions. Substituting these expressions into the modified Friedmann equation and comparing terms of order $\epsilon^0$, we obtain ${C^2}=4\Omega_\Lambda$. Choosing the branch with negative $U$, we have $C=2\sqrt{\Omega_\Lambda}$, and therefore
\begin{equation}
U(N)=-\frac{2\sqrt{\Omega_\Lambda}}{\sqrt{\gamma}}+\mathcal O(1)\,.
\label{eq:app_U_asymptotic}
\end{equation}
This scaling explains how the nonlocal contribution to the Friedmann equation can remain finite as $\gamma\rightarrow0$, despite the divergence of the auxiliary field itself.

The time-dependent corrections follow from the auxiliary-field equations. At leading order, the equation for $U$ gives
\begin{equation}
u''+(3+\xi_\Lambda)u'=6(2+\xi_\Lambda)\,,
\label{eq:app_u_linear}
\end{equation}
where $\xi_\Lambda\equiv E'/E$. Similarly, the equation for $V$ becomes
\begin{equation}
q''+(3+\xi_\Lambda)q'=\frac{1}{E^2}\,.
\label{eq:app_q_linear}
\end{equation}
These equations are linear and depend only on the reference $\Lambda$CDM background. In accordance with the initial conditions adopted in the main text, we impose
\begin{equation}
u(N_{\rm ini})=u'(N_{\rm ini})=0\,,
\qquad
q(N_{\rm ini})=q'(N_{\rm ini})=0\,.
\label{eq:app_linear_initial_conditions}
\end{equation}
To determine the correction $d(N)$, we expand the modified Friedmann equation to first order in $\epsilon$. This yields
\begin{equation}
d(N)=-\frac{C}{2}\left[b+u(N)\right]-CE^2(N)Q(N)\,,
\label{eq:app_d_intermediate}
\end{equation}
where
\begin{equation}
Q(N)\equiv3q(N)+3q'(N)-\frac{1}{2}q'(N)u'(N)\,.
\label{eq:app_Q_definition}
\end{equation}
The normalization condition $h(0)=1$ implies $d(0)=0$, which fixes $b=-u(0)-2Q(0)$. Consequently,
\begin{equation}
d(N)=-C\left[\frac{u(N)-u(0)}{2}+E^2(N)Q(N)-Q(0)\right]\,.
\label{eq:app_d_solution}
\end{equation}
Equation~\eqref{eq:app_d_solution} determines the leading departure from $\Lambda$CDM entirely in terms of the solutions of Eqs.~\eqref{eq:app_u_linear} and \eqref{eq:app_q_linear}. The first correction to the expansion history can therefore be obtained without solving the full nonlinear system at finite $\gamma$.

%%%%%%%%%%%%%%%%%%%%%%%%%%%%%%%%%%%%%%%%%%%%%%%%%%%%%
\subsection{Determination of the CPL consistency relation}
\label{app:CPL_asymptotic_slope}
%%%%%%%%%%%%%%%%%%%%%%%%%%%%%%%%%%%%%%%%%%%%%%%%%%%%%

We now relate the asymptotic expansion to the effective CPL parameters. To this end, we define the contribution of the nonlocal sector to the normalized expansion rate as
\begin{equation}
D(N)\equiv h^2(N)-\Omega_m^0e^{-3N}-\Omega_r^0e^{-4N}\,.
\label{eq:app_D_definition}
\end{equation}
Using Eq.~\eqref{eq:app_Hubble_expansion}, we have
\begin{equation}
D(N)=\Omega_\Lambda+\epsilon d(N)
+\mathcal O(\epsilon^2)\,.
\label{eq:app_D_expansion}
\end{equation}
The effective conservation equation implies
\begin{equation}
w_{\rm DE}=-1-\frac{1}{3}\frac{D'}{D}\,.
\label{eq:app_wDE_D}
\end{equation}
Evaluating this expression and its first derivative at $N=0$ gives
\begin{equation}
1+w_0=-\frac{D_0'}{3D_0}\,,
\qquad
w_a=\frac{1}{3}\left[
\frac{D_0''}{D_0}
-\left(\frac{D_0'}{D_0}\right)^2
\right]\,.
\label{eq:app_CPL_D_exact}
\end{equation}
These expressions are equivalent to the local mapping derived in the previous subsection, but are particularly convenient for studying the $\Lambda$CDM-like limit. Substituting Eq.~\eqref{eq:app_D_expansion}, we obtain
\begin{equation}
1+w_0=-\frac{\epsilon}{3\Omega_\Lambda}d'(0)
+\mathcal O(\epsilon^2)\,,
\qquad
w_a=\frac{\epsilon}{3\Omega_\Lambda}d''(0)
+\mathcal O(\epsilon^2)\,.
\label{eq:app_CPL_asymptotic}
\end{equation}
Eliminating $\epsilon$ yields the local consistency relation
\begin{equation}
w_a=-\alpha_{\rm NL}(1+w_0)
+\mathcal O\!\left((1+w_0)^2\right)\,,
\label{eq:app_consistency_relation}
\end{equation}
where
\begin{equation}
\alpha_{\rm NL}=\lim_{\gamma\rightarrow0}
\left(-\frac{w_a}{1+w_0}\right)
=\frac{d''(0)}{d'(0)}
\,.
\label{eq:app_alpha_asymptotic}
\end{equation}
The result assumes $d'(0)\neq0$ and identifies $\alpha_{\rm NL}$ with the tangent of the path-B trajectory at its $\Lambda$CDM-like endpoint. Unlike the existence of a local linear relation, which follows generically from the regularity of a one-parameter curve, the value of $\alpha_{\rm NL}$ is determined by the underlying nonlocal dynamics.

The derivatives entering Eq.~\eqref{eq:app_alpha_asymptotic} follow directly from Eq.~\eqref{eq:app_d_solution},
\begin{equation}
d'(0)=-C\left[
\frac{u'_0}{2}
+(E^2)'_0Q_0+Q'_0
\right]\,,
\label{eq:app_d_prime}
\end{equation}
and
\begin{equation}
d''(0)=-C\left[
\frac{u''_0}{2}
+(E^2)''_0Q_0
+2(E^2)'_0Q'_0+Q''_0
\right]\,.
\label{eq:app_d_double_prime}
\end{equation}
The derivatives of $Q$ can be evaluated using Eq.~\eqref{eq:app_Q_definition}, while those of $u$ and $q$ follow from the linear equations of motion. For the fiducial cosmological parameters adopted in Sec.~III, namely $\Omega_m^0=0.3114$, $h=0.6751$, and $\Omega_r^0=4.15\times10^{-5}h^{-2}$, we integrate Eqs.~\eqref{eq:app_u_linear} and \eqref{eq:app_q_linear} from $N_{\rm ini}=-20.7233$. The resulting present-day values are
\begin{equation}
u(0)\simeq16.0783\,,
\qquad
u'(0)\simeq3.1079\,,
\qquad
q(0)\simeq0.13907\,,
\qquad
q'(0)\simeq0.26833\,.
\label{eq:app_linear_numerical}
\end{equation}
Evaluating the remaining derivatives through the linear equations, we obtain $d'(0)\simeq-3.13975$ and $d''(0)\simeq-2.15366$ and hence
\begin{equation}
\alpha_{\rm NL}\simeq0.68593\,.
\label{eq:app_alpha_numerical}
\end{equation}
More explicitly, the leading-order expressions for the effective CPL parameters read
\begin{equation}
w_0=-1+1.5201\sqrt{\gamma}
+\mathcal O(\gamma)\,,
\qquad
w_a=-1.0427\sqrt{\gamma}
+\mathcal O(\gamma)\,.
\label{eq:app_CPL_gamma_expansion}
\end{equation}
The corresponding consistency relation is therefore
\begin{equation}
w_a=-0.68593(1+w_0)
+\mathcal O\!\left((1+w_0)^2\right)\,.
\label{eq:app_CPL_final_relation}
\end{equation}
The coefficient obtained here is consistent with the ratios $-w_a/(1+w_0)\simeq0.69$ and $0.75$ found for the representative finite-$\gamma$ solutions discussed in Sec.~\ref{sec:numsol}. The asymptotic calculation provides a more precise determination of the limiting tangent and explains the approximately linear orientation of the theoretical trajectory displayed in Fig.~\ref{gammaVSw}.

We emphasize that the numerical value of $\alpha_{\rm NL}$ depends on the standard cosmological parameters through the reference $\Lambda$CDM background and the corresponding linearized auxiliary-field evolution. The derivation is based on an asymptotic expansion of the path-B solutions in powers of $\sqrt{\gamma}$, and its leading-order prediction applies sufficiently close to the $\Lambda$CDM-like endpoint.

%%%%%%%%%%%%%%%%%%%%%%%%%%%%%%%%%%%%%%%%%%%%%%%%%%%%%%%%%%%%%%%%%%%%%%%
\bibliographystyle{apsrev}
\bibliography{amendolamodnonloc,Henrik,nonlocalRefsTSK,ayuso}
%%%%%%%%%%%%%%%%%%%%%%%%%%%%%%%%%%%%%%%%%%%%%%%%%%%%%%%%%%%%%%%%%%%%%%%

\end{document}